\documentclass[a4paper,fleqn,usenatbib]{mnras}

\usepackage{mathptmx}

\usepackage[T1]{fontenc}
\usepackage{ae,aecompl}

\usepackage{graphicx}	
\usepackage{amsmath}	
\usepackage{amssymb}	
\usepackage{ifxetex}

\def\SuAB{S$\underline{\rm A}$B}

\def\urs{$\underline{\rm r}$s}
\def\rus{r$\underline{\rm s}$}
\def\SAuB{SA$\underline{\rm B}$}

\def\url{$\underline{\rm r}$l}
\def\rul{r$\underline{\rm l}$}
\def\u0a{$\underline{\rm 0}$/a}
\def\0ua{0/$\underline{\rm a}$}
\def\uab{$\underline{\rm a}$b}
\def\aub{a$\underline{\rm b}$}
\def\ubc{$\underline{\rm b}$c}
\def\buc{b$\underline{\rm c}$}
\def\ucd{$\underline{\rm c}$d}
\def\cud{c$\underline{\rm d}$}
\def\udm{$\underline{\rm d}$m}

\def\irpl{r$^{\prime}$l}

\def\Rprime{R$^{\prime}$}
\def\Rone{R$_1$}

\def\RoneP{R$_1^{\prime}$}
\def\RtwoP{R$_2^{\prime}$}
\def\RoneRtwoP{R$_1$R$_2^{\prime}$}
\def\RonePRtwoP{R$_1^{\prime}$R$_2^{\prime}$}
\def\RoneRtwo{R$_1$R$_2$}
\def\RonetwoP{R$_{12}^{\prime}$}
\def\RprimeL{R$^{\prime}$L}
\def\RonePL{R$_1^{\prime}$L}
\def\RoneL{R$_1$L}
\def\cRuL{(R$\underline{\rm{L}}$)}

\title[The Systematics of CVRHS Galaxy Morphology]{The Systematics of Galaxy Morphology in the Comprehensive
de Vaucouleurs revised Hubble-Sandage Classification System: Application to the EFIGI Sample}

\author[R. J. Buta]{
Ronald J. Buta\thanks{E-mail: rbuta@ua.edu}
\\
Department of Physics \& Astronomy,University of Alabama, Box
870324, Tuscaloosa, AL 35487
}
\date{Accepted XXX. Received YYY; in original form ZZZ}

\pubyear{2018}

\begin{document}
\label{firstpage}
\pagerange{\pageref{firstpage}--\pageref{lastpage}}
\maketitle

\begin{abstract}
This paper is the third which examines galaxy morphology from the point
of view of comprehensive de Vaucouleurs revised Hubble-Sandage (CVRHS)
classification, a variation on the original de Vaucouleurs
classification volume that accounts for finer details of galactic
structure, including lenses, nuclear structures, embedded disks, boxy
and disky components, and other features. The classification is applied
to the EFIGI sample, a well-defined set of nearby galaxies which were
previously examined by Baillard et al. and de Lapparent et al. The
survey is focussed on statistics of features, and brings attention to
exceptional examples of some morphologies, such as skewed bars, blue
bar ansae, bar-outer pseudoring misalignment, extremely elongated inner
SB rings, outer rings and lenses, and other features that are likely
relevant to galactic secular evolution and internal dynamics. The
possibility of using these classifications as a training set for
automated classification algorithms is also discussed.
\end{abstract}



\section{Introduction}

Comprehensive galaxy morphology and classification refers to standard
galaxy classification with more emphasis than usual on the fine details
of galactic structure. Such an approach to morphology is warranted by
theoretical progress in understanding galactic structure and evolution
(Kormendy 2012), advances in the sophistication of numerical simulation
models of galaxies (e.g., Dickinson et al. 2018; Eliche-Moral et al.
2018), and by the explosion of high quality, multi-wavelength images of
nearby and very distant galaxies already available or coming in the
near future (Dom\'inguez S\'anchez et al. 2019). After nearly a
century since Hubble (1926) first published his ideas on galaxy
morphology, galaxy classification is still an essential step in the
study of the basic properties of galaxies and of cosmology and the
structure of the Universe. As noted by Simmons et al. (2017), ``Visual
morphologies remain among the most nuanced and powerful measures of
galaxy structure."

The large amount of high quality imaging available at this time makes
it possible to take galaxy morphology and classification to realms it
has not been taken before. Prior to the Sloan Digital Sky Survey (SDSS;
Gunn et al. 1998, 2006; York et al. 2000), few galaxies had large-scale
photographic plate images available for detailed morphological
examination (Sandage \& Bedke 1994). The majority of nearby
galaxies had their morphological classifications judged instead from
small-scale sky survey prints or plates (e.g., the Palomar Sky Survey,
the ESO-$B$ sky survey, and the SRC-$J$ sky survey; de Vaucouleurs et
al. 1991, hereafter RC3). The high quality of these surveys provided a
rich source of morphological information, but the small-scale made it
difficult to reliably classify some galaxies, especially early-type
galaxies, and the details in the centers of many galaxies were lost to
either overexposure or the contrast limits of photographic prints.

The SDSS provides enough new high quality image material to allow
classification details to be seen in several hundred thousand
galaxies.  While an inventory of the types of these galaxies would be
important for cosmological studies, the sheer number of objects makes
visual classification by professional astronomers impractical. This led
to the Galaxy Zoo project (Lintott et al. 2008), which drew volunteers
from the general public to classify in a rudimentary way more than
300,000 SDSS galaxies (Willett et al.  2013).  More recently, the
Galaxy Zoo approach has been applied to more distant galaxies,
including 48,000 galaxies in the CANDELS survey (Simmons et al. 2017)
and 120,000 galaxies in archival {\it Hubble Space Telescope} imaging
(Willett et al. 2017).

Several large professional morphological surveys of SDSS galaxies have
been made. For example, Fukugita et al. (2007) classified 2253 SDSS
galaxies in a simplified revised Hubble system; their final types are
based on the independent classifications of three astronomers. Nair \&
Abraham (2010) used a more comprehensive approach, classifying 14,034
galaxies having 0.01 $<$  $z$ $<$ 0.1 in a system similar to RC3 and to
the {\it Revised Shapley-Ames} catalog (Sandage and Tammann 1981;
Sandage and Bedke 1994), and including additional features like lenses,
tidal tails, and warps. 

One of the largest efforts of multiple astronomers was made by Baillard
et al. (2011), where a team of 10 professional astronomers classified
subsets of 4458 RC3 galaxies for a survey titled ``Extraction of the
Idealized Forms of Galaxies in Images" (French acronym:  EFIGI). The
EFIGI sample was chosen to have many examples of all galaxy types, and
the use of multiple classifiers allowed checks on personal
classification equations and homogenization of the database.  More
recently, Ann et al. (2015) presented visual classifications of 5836
galaxies having $z$ $<$ 0.01. This sample includes many dwarf galaxies
and complements other sources, such as Nair and Abraham (2010).

Modern morphological studies do not always involve application of
classification symbolism as in the Hubble or de Vaucouleurs atlases of
galaxies (e.g., classifications like SBb, SA(rs)a, etc.; Sandage 1961;
Buta et al.  2007). Fukugita et al. (2007) only present modified
numerical $T$-types and no information on other morphological features
of their sample galaxies. Nair \& Abraham (2010) use an RC3 coded
approach for $T$-types, but for other features (bars, rings, lenses,
etc.), a code based on a sum of powers of 2 was used rather than
conventional letter symbolism.  Baillard et al. (2011) classified the
EFIGI sample using numerical codings of 16 visual ``attributes",
including apparent bulge-to-total luminosity ratio, properties of
spiral arms, bars, rings, dust features, star-forming regions, and
environmental characteristics. With these numerical codings and 10
classifiers, Baillard et al. were able to define an ``EFIGI Hubble
sequence" (EHS) and carry out a ``morphometric" analysis of their
sample. In each of these cases, the goal was not only to provide
reliable morphological information on large numbers of nearby galaxies,
but also to set up standards for facilitating automatic classification
algorithms.

This paper is focussed on a re-examination of the EFIGI sample using
the Comprehensive de Vaucouleurs revised Hubble-Sandage (CVRHS)
classification system, in order to examine the systematics of
morphology within the system. The CVRHS is a variation on the de
Vaucouleurs (1959) classification volume that recognizes features of
interest beyond the original system.  The goals of the reclassification
are: (1) to provide new classifications of EFIGI galaxies of a similar
nature to RC3 classifications, but which supercede the latter and are
useful for statistical studies; (2) to identify unusual or special
objects that may shed light on galactic evolutionary processes; and (3)
to help define the CVRHS as a point of view for further studies of
galaxy morphology, including the facilitation of automated galaxy
classification and its application to large samples of galaxies.

The CVRHS is briefly described in sections 2 and 3. The classification
procedure is described in section 4, and both an internal and external
comparison of classifications is described in section 5. The mean
catalogue is described in section 6.  Some statistics in the catalogue
are examined in section 7. Finally, montages of special features of
interest are provided in section 8.

\section{CVRHS Morphological Classification}

The CVRHS is a galaxy classification system tied to that of Hubble
(1926; see also Sandage 1961) but as revised by de Vaucouleurs (1959),
the latter also called the de Vaucouleurs revised Hubble-Sandage or
VRHS system, which is the system used in RC3. Because of extensive
continuing use of RC3 nearly 30 years after its publication, the VRHS
is the ``most applied" classification system in extragalactic studies,
at least for nearby galaxies.  The comprehensive element was added to
the VRHS in several studies, but is most thoroughly described by Buta
et al. (2015).

The CVRHS system has been most recently applied to two samples:  the
nearly 2400 galaxies in the {\it Spitzer} Survey of Stellar Structure
in Galaxies (S$^4$G; Sheth et al.  2010) by Buta et al.  (2015), and to
3962 ringed galaxies drawn from the Galaxy Zoo 2 project (Willett et
al. 2013; Buta 2017a). The historical waveband for galaxy
classification is a blue-sensitive photographic emulsion, which neither
of these studies used. The S$^4$G classification used 3.6$\mu$m
mid-infrared images in units of magnitudes per square arcsecond, while
the ringed galaxy classification used SDSS color images (Lupton et al.
2004). For the EFIGI classification, SDSS $g$-band images (effective
wavelength 477 nm) in units of magnitudes per square arcsecond are
used. This also does not match the historical waveband for galaxy
classification, but nevertheless is one of the closest modern
approximations. Eskridge et al. (2002) discuss the systematic
differences that can arise in galaxy classification when the old blue
light systems are applied at a drastically different wavelength
such as the infrared.

\section{Scientific Advantages of CVRHS Classification}

It is a fair question to ask what scientific advantages the CVRHS
system might have over other approaches to galaxy classification.
Consistent with the VRHS, the CVRHS provides more detailed
morphological information than conventional Hubble types (e.g., Sandage
1961) without being too unwieldy. For most galaxies, a CVRHS type is
little different from a VRHS type. Nevertheless, the detailed nature of
CVRHS classification is ideal for interpreting SDSS images of
relatively nearby galaxies, which is important because modern
simulations (e.g., Illustris; Dickinson et al. 2018) have reached the
point where model galaxies resemble real galaxies well enough that they
can be compared to specific SDSS galaxies.

One advantage of CVRHS classification over others is the extent to
which inner, outer, and nuclear varieties are recognized, These
characteristics of galaxy morphology are important aspects of internal
disk galaxy dynamics. For example, a significant part of CVRHS
morphology involves the recognition of the different types of ring
phenomena, including pseudorings and lenses, which have been tied to
galactic secular evolutionary processes (Kormendy 1979; 2012).
Although inner and outer rings are already recognized in the VRHS
system, CVRHS morphology in addition recognizes nuclear rings, special
subclasses of outer rings, as well as rare multiple inner, outer, and
nuclear rings, pseudorings, and lenses. The \Rone, \RoneP, \RtwoP, and
\RoneRtwoP\ subclasses of outer rings and pseudorings (Buta and Crocker
1991) are an example of CVRHS features that can be tied to specific
aspects of internal dynamics, such as resonant (Buta and Combes 1996)
or manifold (Athanassoula et al. 2010 and references therein)
dynamics.

It is also fair to ask whether and how CVRHS classification facilitates
the identification of rare or special objects.  While special cases
could of course be noted without appealing to CVRHS morphology, the
CVRHS provides a better context for recognizing such cases because the
level of detail demands a close inspection of the features defining the
classification. Also, when applying the CVRHS to a sample, every galaxy
is viewed in the context of the whole, which makes special cases stand
out. Finally, any classification allows the drawing of samples for further 
study, and the more detailed the classification, the more specific a sample
can be.

\section{CVRHS Classification of EFIGI Galaxies}

The EFIGI sample was chosen for this study because the sample was
carefully selected  by Baillard et al. (2011) to include mainly
galaxies having a reliable RC3 classification; the sample is large
enough to have many examples of each galaxy type, but small enough to
allow CVRHS classification by a single person in a reasonable amount of
time; and the authors posted on a public website
(www.astromatic.net/projects/efigi) the full set of color images and
individual filter images they used for their study, making it possible
to focus entirely on morphology and not on image preparation.  De
Lapparent et al. (2011) summarize other statistics of the EFIGI
catalogue. Redshifts range from nearly 0 to 0.07, absolute $g$-band
magnitudes range from $-$13 to $-23$, and linear diameters range from
1-100 kpc.

The CVRHS classification of the EFIGI sample is based on logarithmic,
background-subtracted SDSS images converted to units of magnitudes per
square arcsecond. This is the display system of the de Vaucouleurs
Atlas of Galaxies (deVA, Buta et al. 2007). The images are from Data
Release 4 (Adelman-McCarthy et al. 2006), and all such images have
known zero points in the $AB$ magnitude system. The typical zero point
in the $g$-band is 26.5 mag arcsec$^{-2}$, which allowed a homogeneous
display of all of the galaxies. The conversion of units to mag
arcsec$^{-2}$ is essential to CVRHS morphology as applied in the deVA,
although an actual zero point is not a stringent requirement.

The classification of the EFIGI galaxies was carried out in three
phases: Phase 1 (2012) and Phase 2 (2017) involved independent
classifications of the full sample of 4458 galaxies, while phase 3
(2018) involved only 2019 spirals approximately in the stage range S0/a to
Sd. The reason for such an approach is that multiple, independent
examinations of the database of images allow internal consistency
checks of the morphological interpretations. This is important because
an observer may not view a galaxy in exactly the same manner from
one examination to another, and combining multiple phases can improve
the final classification.

The multi-phase approach to galaxy classification is an important part
of this study, and its successful application depends on how well the
different phases can be combined. For this purpose, several steps were
used: (1) averaging the stage (E, S0, S, I); (2) averaging the family
(SA, SAB, SB); (3) averaging the main part of the inner variety (r, rs,
s, rl, l); (4) selecting the remaining parts of the inner variety; and
(5) selecting the outer variety. For stage, family, and inner variety,
the phases were combined using a numerical coding system. For stage,
the coding system was the familiar $T$-index used in RC3, which ranges
from $T$=$-$5 for E galaxies to $T$=10 for Im galaxies.  For family, a
modified coding scheme was used:  $F$=0 for SA galaxies, 1
for \SuAB\ galaxies, 2 for SAB galaxies, 3 for \SAuB\ galaxies, and 4
for SB galaxies. For inner variety, $IV$=0 for (s), 1 for (\rus ), 2
for (rs). 3 for \urs ), 4 for (r), 5 for (\url), 6 for (rl), 7 for
(\rul ), and 8 for (l). The underline categories \SuAB, \SAuB, (\urs ),
(\rus ) (de Vaucouleurs 1963) are directly applied in family and inner
variety classifications in each phase.  However, underline stages
(e.g., S\uab, S\buc, etc.) appear only in the averaged classifications.

\section{Comparison of Classifications}

\subsection{Internal Consistency}

\begin{figure*}
\includegraphics[width=\textwidth]{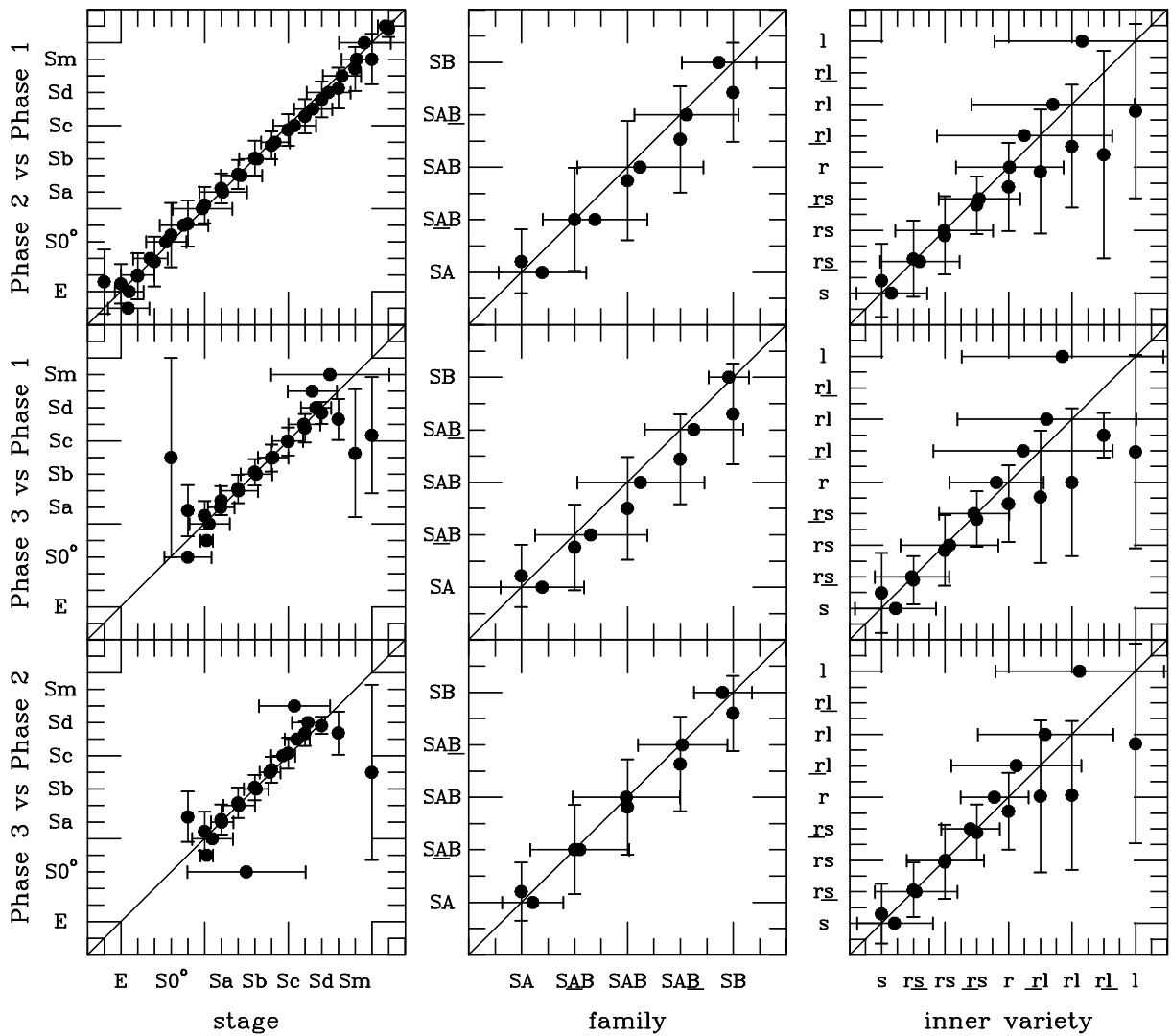}
\vspace{-10truemm}
\caption{Comparison of stage, family, and inner variety classifications between Phases 1, 2, and 3
}
\label{fig:intFTV}
\end{figure*}

Figure~\ref{fig:intFTV} shows how well the stage, family, and inner
variety classifications from the three phases agree. Each stage graph
plots two sets of points:  $<T_j>$ versus $T_i$ and $T_j$ versus
$<T_i>$, for phases $i$ and $j$; each family graph plots $<F_j>$ versus
$F_i$ and $F_j$ versus $<F_i>$; and each inner variety graph plots
$<IV_j>$ versus $IV_i$ and $IV_j$ versus $<IV_i>$. The error bars in
each direction are 1$\sigma$ standard deviations. In general, the
agreement between stage, family, and inner variety classifications from
the three phases is good. Table~\ref{tab:comp12} shows that of 2019
galaxies having three independent classifications, 120 (or 5.9\%)
received an identical full classification in all three phases, while
out of 4458 galaxies, 952 (21.4\%) received two identical full
classifications. For more than half the sample, the same stage, family,
or variety was assigned ($|\Delta T|$, $|\Delta F|$, or $|\Delta (IV)|$
=0). A small percentage have $|\Delta T|$, $|\Delta F|$, or $|\Delta
(IV)|$ $>$3; such large discrepancies are often associated with complex
objects difficult to fit into the classification system.

Table~\ref{tab:fourway} summarizes the results of an analysis of the
internal dispersions in the assignments of CVRHS stage, family, and
inner variety classifications for EFIGI galaxies. In general,
systematic differences in these attributes for the different phases are
small. The dispersions are derived from

\begin{equation} \label{eq:sigmat}
\sigma_{ij}^2 = {1\over N}\Sigma (T_i-T_j)^2 
\end{equation}

\noindent
for phases $i$ and $j$, where $N$ is the total number of galaxies in
the comparison. From these combined dispersions, the individual
$\sigma_i$ can be derived using linear least squares. The results are
given in terms of intervals: for example, 1 stage interval is a
difference such as Sab to Sb; 1 family interval is a difference such as
SA to S$\underline{\rm A}$B; and 1 inner variety interval is a
difference such as ($\underline{\rm r}$s) to (r). The analysis is
restricted to spirals because only spirals in the range S0/a to Sd
(based on an initial average of phase 1 and 2 classifications) have
classifications in all three phases. There is an indication that Phase
2 and 3 classifications are more internally consistent than are phase 1
classifications. For example, $\sigma_1(F)$ = 0.76 while $\sigma_2(F)$
= 0.52 and $\sigma_3(F)$ = 0.57. Such differences are not likely to be
significant, and Table~\ref{tab:fourway} adopts the averages
$<\sigma_i(T)>$ = 0.7 stage intervals, $<\sigma_i(F)$> = 0.6 family
intervals, and $<\sigma_i(IV)>$ = 1.1 inner variety intervals.
These dispersions are characteristic of the classifications for a
single phase, but if $n$ phases are averaged, then the internal
dispersion is derived as $<\sigma_i>$/$\sqrt{n}$.

\begin{table}
\centering
\caption{Internal comparison of classifications}
\label{tab:comp12}
\begin{tabular}{lrrrrrr}
\hline
Comparison & $n_{12}$ & \% $N$ & $n_{13}$ & \% $N$ & $n_{23}$ & \% $N$ \\
\hline
3 identical    &   120 &   5.9 &         &       &       &      \\
$N$            &  2019 &       &         &       &       &      \\
2 identical    &   952 &  21.4 &         &       &       &      \\
$N$            &  4458 &       &         &       &       &      \\
                   &     &      \\
$\Delta T$=0   &  2283 &  52.5 &  1119 &  56.3 &  1179 &  59.3 \\
$|\Delta T|$=1 &  1437 &  33.0 &   716 &  36.0 &   679 &  34.2 \\
$|\Delta T|$=2 &   405 &   9.3 &   111 &   5.6 &   104 &   5.2 \\
$|\Delta T|$=3 &   145 &   3.3 &    26 &   1.3 &    12 &   0.6 \\
$|\Delta T|>$3 &    81 &   1.9 &    17 &   0.9 &    13 &   0.7 \\
$N$            &  4351 &       &  1989 &       &  1987 &       \\
                   &     &      \\
$\Delta F$=0   &  1977 &  65.2 &  1249 &  63.0 &  1365 &  68.8 \\
$|\Delta F|$=1 &   568 &  18.7 &   452 &  22.8 &   446 &  22.5 \\
$|\Delta F|$=2 &   442 &  14.6 &   254 &  12.8 &   162 &   8.2 \\
$|\Delta F|$=3 &    25 &   0.8 &    19 &   1.0 &     8 &   0.4 \\
$|\Delta F|>$3 &    21 &   0.7 &     9 &   0.5 &     2 &   0.1 \\
$N$            &  3033 &       &  1983 &       &  1983 &       \\
                   &     &      \\
$\Delta IV$=0   &  1751 &  70.4 &  1219 &  68.5 &  1252 &  67.6 \\
$|\Delta IV|$=1 &   395 &  15.9 &   321 &  18.0 &   371 &  20.0 \\
$|\Delta IV|$=2 &   316 &  12.7 &   219 &  12.3 &   212 &  11.4 \\
$|\Delta IV|$=3 &    21 &   0.8 &    19 &   1.1 &    14 &   0.8 \\
$|\Delta IV|>$3 &     3 &   0.1 &     2 &   0.1 &     4 &   0.2 \\
$N$            &  2486 &       &  1780 &       &  1853 &       \\
                   &     &     \\
\hline
\end{tabular}
\end{table}

\begin{table}
\renewcommand\thetable{2}
\centering
\caption{Internal Agreement between stage and family classifications in
Phases 1, 2, and 3. Each column gives the rms dispersion between the
two phases ($i$ and $j$) indicated. The numbers in parentheses next to
each value are the mean difference ($<T_i-T_j>$, $<F_i-F_j>$, or
$<IV_i-IV_j>$) and  the number of galaxies in the comparison. The
individual $\sigma_i$ are derived from a linear least squares
analysis.
}
\label{tab:fourway}
\begin{tabular}{cccc}
\hline
 $i\rightarrow$ & 1 & 2 & 3  \\
 $j$ &  &  & \\
 $\downarrow$ & & & \\
\hline
       &             &  Stage       &              \\
2 &   1.21(0.15,1984) &  .......... & .......... \\
3 &   1.02($-$0.01,1989) &   0.93($-$0.16,1987) & .......... \\
       &             &             &              \\
$\sigma_i(T)$   & 0.90 & 0.80 & 0.47  \\
$<\sigma_i(T)>$ & {\bf 0.7} &      &       \\
       &             &             &              \\
       &             & Family      &              \\
2 &   0.92(0.20,1973) &  .......... & .......... \\
3 &   0.95(0.26,1983) &   0.78(0.06,1983) & .......... \\
       &             &             &              \\
$\sigma_i(F)$   & 0.76 & 0.52 & 0.57  \\
$<\sigma_i(F)>$ & {\bf 0.6} &      &       \\
       &             &             &              \\
       &             & Inner Variety      &              \\
2 &   1.57(0.05,1894) &  .......... & .......... \\
3 &   1.59(0.10,1905) &   1.36(0.06,1933) & .......... \\
       &             &             &              \\
$\sigma_i(IV)$   & 1.25 & 0.95 & 0.98  \\
$<\sigma_i(V)>$ & {\bf 1.1} &      &       \\
\hline
\end{tabular}
\end{table}

\subsection{External Comparisons}

\begin{figure}
\includegraphics[width=\columnwidth]{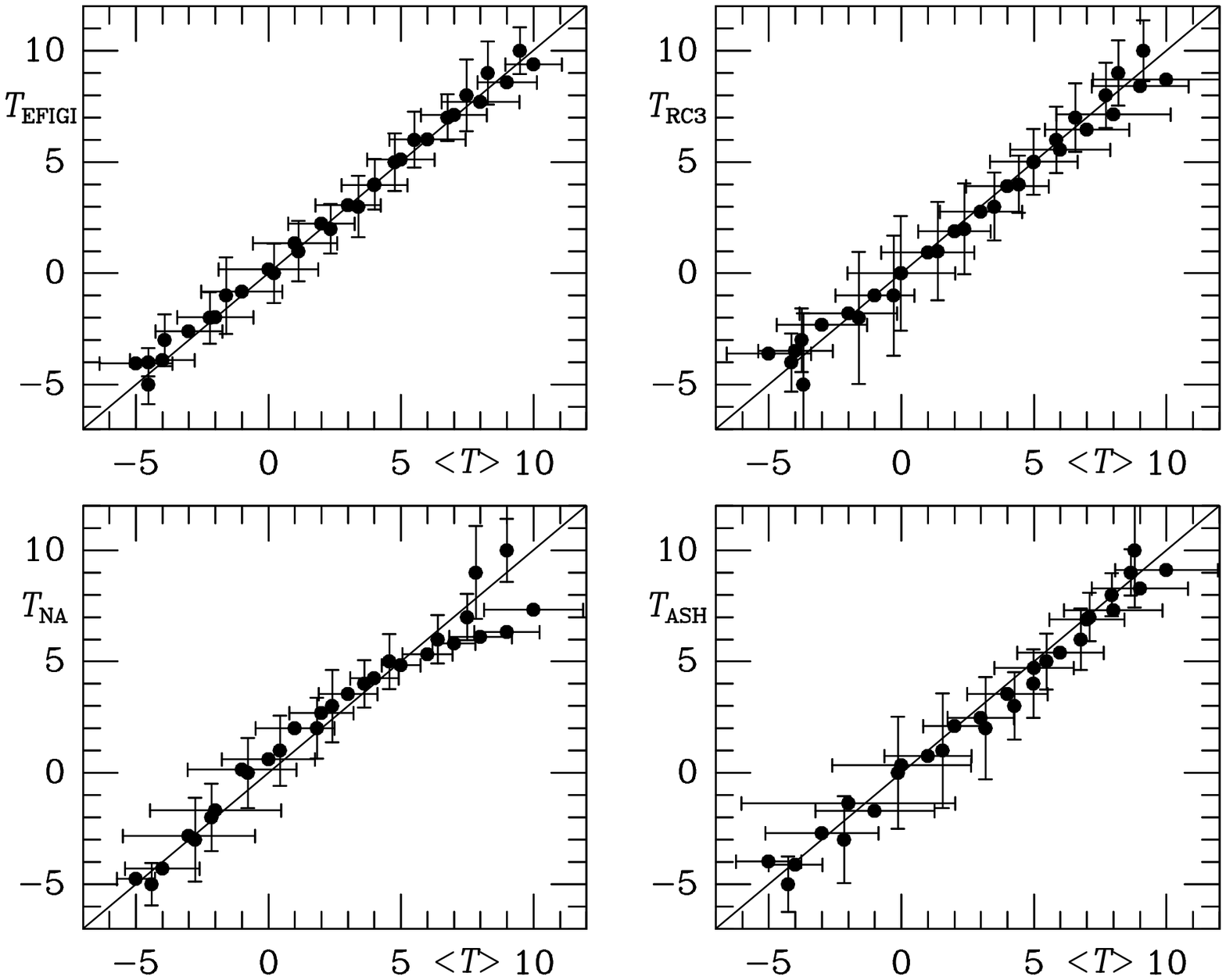}
\vspace{8truemm}
\caption{Comparison of Table 1 mean stage classifications with stage
classifications from other sources: $T_{EFIGI}$ from Baillard et al.
(2011; 4386 galaxies), $T_{RC3}$ from de Vaucouleurs et al.  (1991;
4420 galaxies), $T_{NA}$ from Nair \& Abraham (2010; 1389 galaxies),
and $T_{ASH}$ from Ann et al. (2015; 1250 galaxies.)
}
\label{fig:extstages}
\end{figure}

\begin{figure}
\includegraphics[width=\columnwidth]{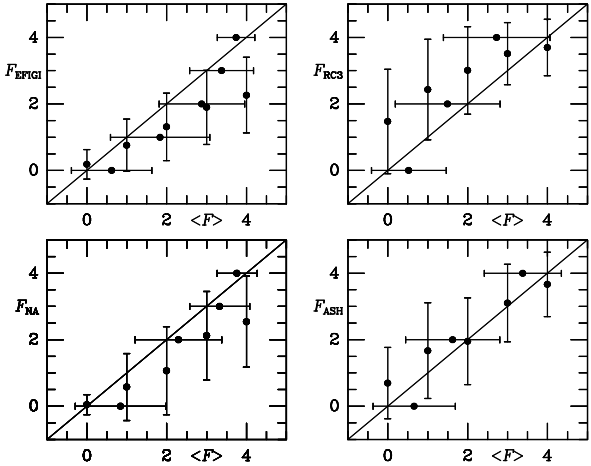}
\vspace{8truemm}
\caption{Comparison of Table 1 mean family classifications
with family classifications from other sources: $F_{EFIGI}$ from
Baillard et al. (2011; 3319 galaxies), $F_{RC3}$ from de Vaucouleurs et al.
(1991; 1884 galaxies), $F_{NA}$ from Nair
\& Abraham (2010;1158 galaxies), and $F_{ASH}$ from Ann et al. (2015; 715 galaxies)
}
\label{fig:extfams}
\end{figure}

External comparisons between the CVRHS types and published types from
other sources are presented in Figures ~\ref{fig:extstages} and
~\ref{fig:extfams}. Four external sources are examined:  Baillard et
al. (2011, source "EFIGI"), de Vaucouleurs et al. (1991, source "RC3"),
Nair and Abraham (2010; source "NA"), and Ann et al. (2015,
source "ASH"). Only stage and family classifications are compared (ASH
do not judge inner or outer varieties).  The line in each frame is for
reference only and not a linear fit.

For the purposes of the comparison, mean stages ($<T>$) and mean
families ($<F>$) were derived as unweighted averages of phases 1-3 for
spirals and phases 1-2 for the remaining types. NA family
classifications are specified by numbers $F$ = 2$^i$, where $i$ = 1 for
a strong bar, 2 for an intermediate bar, and 3 for a weak bar, and
$F$=0 for no bar. In Figure~\ref{fig:extfams}, these are translated into
classifications SB, SA$\underline{\rm B}$, SAB, and SA, respectively.
NA noted that their bar classifications mainly emphasized strong
bars.

Figure~\ref{fig:extstages} shows good agreement between $<T>$ and the
EFIGI Hubble sequence $T_{EFIGI}$ for types ranging from $T$ = $-$5 to
$T$ = 10. Agreement with the other sources is good as well. Comparison
between family classifications shows some systematic disagreements
(e.g., $<F>$ is stronger on average than $F_{EFIGI}$ and $F_{NA}$,
but weaker on average than $F_{RC3}$).

This is quantified further in Table~\ref{tab:allway}, again using
equation~\ref{eq:sigmat} and linear least squares. The five sources
[this paper (RB), EFIGI, RC3, NA, and ASH) could not be used together
since there is little or no overlap between NA and ASH. Only the
galaxies in common with all of the available sources are used for this
analysis.  Table~\ref{tab:allway} presents separate analyses of sources
RB, EFIGI, RC3, and NA, and then RB, EFIGI, RC3, and ASH. The average
external dispersion for the modern sources is $<\sigma_e(T)>$ = 1.1
stage intervals, while that for RC3 is $<\sigma_e(T)>$ = 1.5 stage
intervals.  For family, the two sets do not give very consistent
results for sources EFIGI and RC3. The average of the modern sources is
$<\sigma_e(F)>$ = 0.8 family intervals while for RC3 it is
$<\sigma_e(F)>$ = 1.5 family intervals. The results of both the
internal and external comparisons are consistent with Buta et al.
(2015) and Naim et al. (1995).

\begin{table}
\centering
\caption{External Agreement Between Classifications.
Each column gives the rms dispersion between the two
sources ($i$ and $j$) indicated. The number in parentheses next to each
value is the mean difference $<T_i-T_j>$. The individual
$\sigma_i$ are derived from a linear least squares analysis. Sources:
EFIGI=Baillard et al. (2011); NA2010=Nair \& Abraham (2010); RC3=de
Vaucouleurs et al. (1991); ASH = Ann et al. (2015)}
\label{tab:allway}
\begin{tabular}{cccccc}
\hline
Source & RB & EFIGI & RC3 & NA \\
 $i\rightarrow$ & 1 & 2 & 3 & 4  \\
 $j$ &  &  & &  \\
 $\downarrow$ & & & &  \\
\hline
      &             &             &             &              \\
    2 &   1.33(0.24) &  .......... &  .......... & ..........   \\
    3 &   1.71(0.02) &   1.80($-$0.228) &  ..........  & ..........  \\
    4 &   1.40(0.22)   &   1.64($-$0.02) &   1.90(0.20)  & ..........  \\
      &             &             &             &   \\
$\sigma_i(T)$  & 0.80 & 1.07 & 1.48 & 1.19 \\
$N$=1389       &      &      &      &      \\
\hline
Source & RB & EFIGI & RC3 & ASH  \\
$i\rightarrow$ & 1 & 2 & 3 & 4  \\
$j$ &  &  & &  \\
$\downarrow$ & & & &  \\
\hline
      &             &             &             &              \\
    2 &   1.56(0.15) &  .......... &  .......... & ..........   \\
    3 &   1.88($-$0.22) &   2.03($-$0.37) &  ..........  & ..........  \\
    4 &   1.73($-$0.32)   &   1.72($-$0.47) &   1.68($-$0.10)  & ..........  \\
      &             &             &             &   \\
$\sigma_i(T)$  & 1.15 & 1.27 & 1.44 & 1.12 \\
$N$=1217       &      &      &      &      \\
\hline
\hline
Source  & RB & EFIGI & RC3 & NA \\
 $i\rightarrow$ & 1 & 2 & 3 & 4 \\
 $j$ &  &  & &  \\
 $\downarrow$ & & & &  \\
\hline
  &             &             &             &             \\
2 &   0.99($-$0.46) &  .......... &  .......... & ..........  \\
3 &   1.71(0.85) &   1.95(1.30) &  ..........  & ..........  \\
4 &   1.17($-$0.57)   &   0.93($-$0.11) &   2.14($-$1.41)  & ..........  \\
  &             &             &             &   \\
$\sigma_i(F)$  & 0.46 & 0.63 & 1.80 & 1.00 \\
$N$=628       &      &      &      &      \\
\hline
Source & RB & EFIGI & RC3 & ASH  \\
 $i\rightarrow$ & 1 & 2 & 3 & 4  \\
 $j$ &  &  & &  \\
 $\downarrow$ & & & &  \\
\hline
   &             &             &             &         \\
 2 &   1.40($-$0.89) &  .......... &  .......... & .......... \\
 3 &   1.41(0.50) &   1.94(1.39) &  ..........  & ..........  \\
 4 &   1.14(0.19)   &   1.60(1.08) &   1.36($-$0.31)  & ..........   \\
             &             &             &             &   \\
$\sigma_i(F)$   & 0.62 & 1.38 & 1.25 & 0.79 \\
$N$=504       &      &      &      &      \\
\hline
\end{tabular}
\end{table}

\section{Mean Catalogue}

Table~\ref{tab:catalog} presents the mean CVRHS classifications for the
4458 EFIGI sample galaxies (followed by notes for three-quarters of the
sample). The galaxies are listed by Principal Galaxy Catalogue (PGC,
Paturel et al. 1989) number, but most also have an alternate name that
can take precedence over the PGC number. The names in column 1 are
formal ``NED names," i. e., the names adopted in the NASA/IPAC
Extragalactic Database\footnote{The NASA/IPAC Extragalactic Database
(NED) is operated by the Jet Propulsion Laboratory, California
Institute of Technology, under contract with the National Aeronautics
and Space Administration.} and were taken from a cross index provided
on the EFIGI website.  The radial velocity listed in column 3 is also
from this cross index list. Column 4 lists the arm class (Elmegreen and
Elmegreen 1987) for those galaxies where the image resolution and
inclination allowed a judgment to be made. Columns 5 and 6 list
numerical codes for the mean stage and family classifications.  The
$T$-type codes are the same as defined in RC3 with the exception that
$T$=11 is used for dwarf elliptical, S0, and spheroidal galaxies.  This
is consistent with what Baillard et al.  (2011) used for these same
types. The coding for  bar classification:  $F$ = 0, 1, 2, 3, and 4 for
types SA, S$\underline{\rm A}$B, SAB, SA$\underline{\rm B}$, and SB,
respectively, is the same scale that Baillard et al. (2011) used,
multiplied by a factor of 4. Column 7 lists the number of phases used
for the averages.  The final mean letter classification is listed in
column 8.

 \begin{table*}
 \centering
 \setcounter{table}{3}
 \caption{CVRHS Classifications for
 4458 EFIGI Galaxies:
 Col. 1: name adopted in NASA/IPAC Extragalactic
 Database (NED);
 col. 2: number in Principal Galaxy Catalogue
 (Paturel et al. 1989);
 col. 3: heliocentric radial velocity from NED;
 col. 4: arm class (as defined by Elmegreen and
 Elmegreen 1987);
 col. 5: mean stage index on the RC3 scale;
 col. 6: mean family index on the scale used
 in this paper;
 col. 7: number of phases used for $<T>$ and $<F>$;
 col. 8: the average classification using CVRHS
 notation.
 (See Buta et al. 2015 for the meaning of different
 CVRHS symbols;
the full table will be made available online).}
 \label{tab:catalog}
 \begin{tabular}{lrrrrrcl}
 \hline
 NED Name & PGC & $V_{\odot}$ & AC & $<T>$ & $<F>$
  & $n$ & Type \\
    &  & km s$^{-1}$ &  &   &   &  &   \\
 1 & 2  & 3 & 4 & 5 & 6 & 7 & 8  \\
\hline
IC   5381         &   212 & 11231 &    &  2.5$\pm$0.7 & 1.0$\pm$1.1 & 2 & \SuAB $_x$(\urs )\aub\   spw \\
NGC  7814         &   218 &  1050 &    &  1.0$\pm$0.5 &             & 2 & Sa  sp / E(d)5 \\
NGC  7808         &   243 &  8787 &    &  0.0$\pm$0.4 & 0.0$\pm$0.3 & 3 & SA(\url )0/a \\
UGC    17         &   255 &   878 &  4 &  9.0$\pm$0.5 & 2.0$\pm$0.4 & 2 & SAB(s)m \\
MCG $-$02-01-015    &   281 & 11491 &    &  5.0$\pm$4.0 & 3.0$\pm$1.1 & 2 & \SAuB $_a$(s)c  pec \\
\hline
\end{tabular}
\end{table*}

\section{Distribution of Morphologies in the CVRHS-EFIGI Catalogue}

\begin{figure}
\includegraphics[width=\columnwidth]{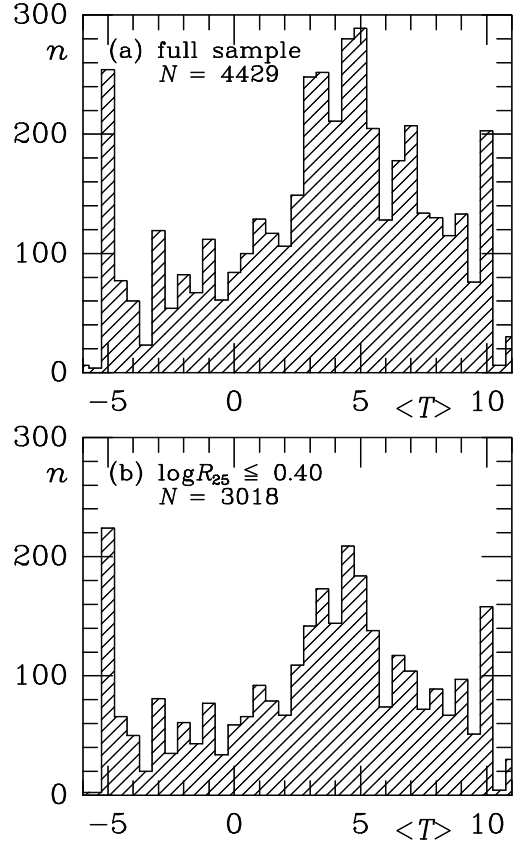}
\vspace{-5truemm}
\caption{Distribution of CVRHS Stages
}
\label{fig:histot}
\end{figure}

\begin{figure}
\includegraphics[width=\columnwidth]{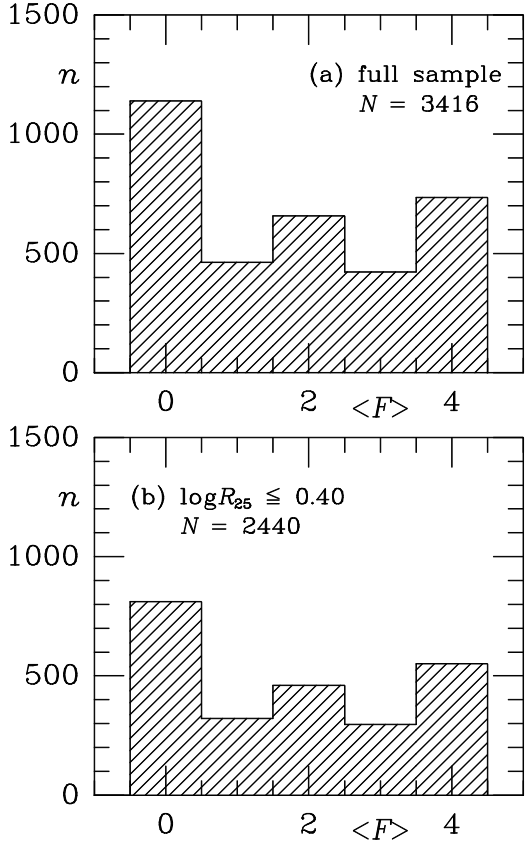}
\vspace{-5truemm}
\caption{Distribution of CVRHS Families
}
\label{fig:histof}
\end{figure}

\begin{figure}
\includegraphics[width=\columnwidth]{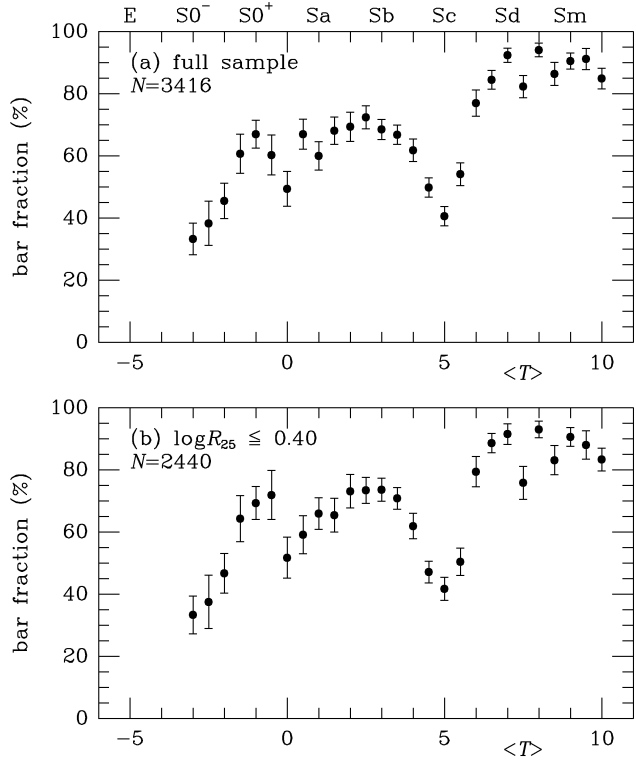}
\vspace{-5truemm}
\caption{Bar fraction as a function of CVRHS stage
}
\label{fig:barfrac}
\end{figure}

\begin{figure}
\includegraphics[width=\columnwidth]{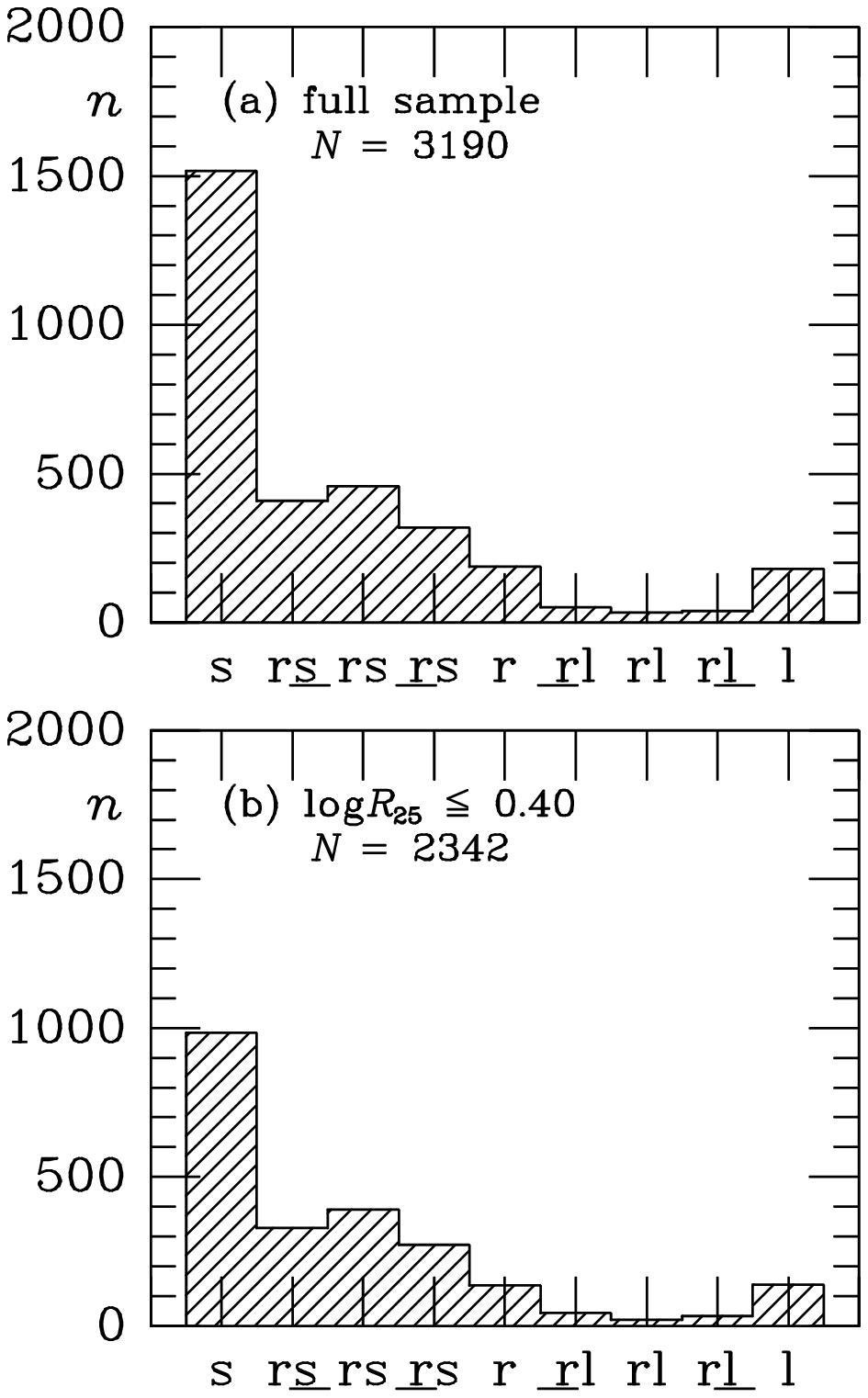}
\vspace{-5truemm}
\caption{Distribution of CVRHS inner varieties
}
\label{fig:histov}
\end{figure}

Figures~\ref{fig:histot}--~\ref{fig:histov} and
Tables~\ref{tab:tstats}-~\ref{tab:ivstats} show distributions of
classifications from the mean catalogue, for two samples in each case:
a full, unrestricted sample, and a second sample restricted to RC3
isophotal axis ratio log$R_{25}$ $\leq$ 0.40. The latter restriction is
meant to exclude galaxies more highly inclined than 66$^o$.  The
distribution of stages is comparable to what Baillard et al. (2011,
their Figure 26) found, except that E and Im galaxies stand out more in
part because the histograms in Figure~\ref{fig:histot} are plotted in
half stage intervals rather than full stage intervals.

In both the full and restricted subsets in Table~\ref{tab:tstats}, Sb
to Sc are the stages with the highest relative frequencies, with
galaxies in the range Sa$\underline{\rm b}$ to S$\underline{\rm c}$d
constituting 36-37\% of the samples. The shape of the distribution of
types is different from that expected for a distance-limited sample,
which would tend to emphasize extreme late-type spirals and magellanic
irregulars (e.g., Buta et al. 2015, their Figure 5; also Buta et al.
1994, their Figure 6). It is also different from a magnitude-limited
sample, which would tend to de-emphasize such galaxies.  Instead, the
distributions are characteristic of an angular diameter-limited sample,
as noted by de Lapparent et al. (2011; see also Buta et al 1994, their
Figure 5).

The statistics of bar classifications shows that family SA has the
highest relative frequency, constituting 33\% of both the full and
restricted samples. This means the EFIGI sample as classified
according to the CVRHS system has a bar fraction of 67\% if the weakest
bars (type S$\underline{\rm A}$B) are included, or 53\% based only on
SAB, SA$\underline{\rm B}$, and SB types.  The bar classifications in
Table~\ref{tab:catalog} are not defined in the same manner as the
Baillard et al. (2011) bar attribute. CVRHS bar classifications account
for bar length, axis ratio, and contrast, while the Baillard et al.
(2011) bar attribute is based mainly on relative bar length.

Figure~\ref{fig:barfrac} and Table~\ref{tab:barstage} show the CVRHS
EFIGI bar fraction as a function of stage. Both the full and restricted
samples show a significant minimum in bar frequency around stage Sc,
with a lesser minimum at stage S0/a. The graphs also show that the bar
fraction for stages Scd and later is greater than 80\%, compared to
$\approx$60\% for stages S0$^+$ to Sb. These trends are similar to what
Buta et al. (2015) found for the S$^4$G sample except that the minimum
in the mid-IR occurs at a slightly earlier stage than Sc. The
implication of this minimum is that conventional Sc galaxies rarely
appear barred. In fact, one of the most common types in the catalogue
is SA(s)c. Buta et al.  (2015) also show that RC3 classifications show
a minimum in bar fraction near stage Sc, although this minimum is much
less significant.

\begin{table}
\centering
\caption{Statistics of Stages
}
\label{tab:tstats}
\begin{tabular}{rrrrrrr}
\hline
$T$      &   $n$    & \%$N$    &   $T$    & $n$      &   \%$N$  \\
1        &     2    & 3        &     1    & 2        &     3    \\
\hline
         &          &          &          &          &          \\
         &          & Full Sample & $N$=4429 &       &          \\
         &          &          &          &          &          \\
    $-$6.0 &        6 &      0.1 &      3.0 &      249 &      5.6 \\
    $-$5.5 &        4 &      0.1 &      3.5 &      252 &      5.7 \\
    $-$5.0 &      254 &      5.7 &      4.0 &      211 &      4.8 \\
    $-$4.5 &       77 &      1.7 &      4.5 &      280 &      6.3 \\
    $-$4.0 &       60 &      1.4 &      5.0 &      289 &      6.5 \\
    $-$3.5 &       23 &      0.5 &      5.5 &      205 &      4.6 \\
    $-$3.0 &      119 &      2.7 &      6.0 &      128 &      2.9 \\
    $-$2.5 &       54 &      1.2 &      6.5 &      178 &      4.0 \\
    $-$2.0 &       82 &      1.9 &      7.0 &      207 &      4.7 \\
    $-$1.5 &       67 &      1.5 &      7.5 &      134 &      3.0 \\
    $-$1.0 &      112 &      2.5 &      8.0 &      130 &      2.9 \\
    $-$0.5 &       61 &      1.4 &      8.5 &      115 &      2.6 \\
     0.0 &       84 &      1.9 &      9.0 &      133 &      3.0 \\
     0.5 &      100 &      2.3 &      9.5 &       76 &      1.7 \\
     1.0 &      129 &      2.9 &     10.0 &      203 &      4.6 \\
     1.5 &      117 &      2.6 &     10.5 &        6 &      0.1 \\
     2.0 &      105 &      2.4 &     11.0 &       30 &      0.7 \\
     2.5 &      149 &      3.4 &     .... &      ... &      ... \\
         &          &          &          &          &          \\
         &          & $logR_{25}$ $\leq$ 0.40 & $N$=3018 &       &          \\
         &          &          &          &          &          \\
    $-$6.0 &        2 &      0.1 &      3.0 &      142 &      4.7 \\
    $-$5.5 &        2 &      0.1 &      3.5 &      173 &      5.7 \\
    $-$5.0 &      224 &      7.4 &      4.0 &      144 &      4.8 \\
    $-$4.5 &       66 &      2.2 &      4.5 &      209 &      6.9 \\
    $-$4.0 &       50 &      1.7 &      5.0 &      184 &      6.1 \\
    $-$3.5 &       20 &      0.7 &      5.5 &      138 &      4.6 \\
    $-$3.0 &       81 &      2.7 &      6.0 &       74 &      2.5 \\
    $-$2.5 &       35 &      1.2 &      6.5 &      117 &      3.9 \\
    $-$2.0 &       61 &      2.0 &      7.0 &      104 &      3.4 \\
    $-$1.5 &       43 &      1.4 &      7.5 &       72 &      2.4 \\
    $-$1.0 &       77 &      2.6 &      8.0 &       89 &      2.9 \\
    $-$0.5 &       34 &      1.1 &      8.5 &       67 &      2.2 \\
     0.0 &       59 &      2.0 &      9.0 &       97 &      3.2 \\
     0.5 &       66 &      2.2 &      9.5 &       51 &      1.7 \\
     1.0 &       92 &      3.0 &     10.0 &      158 &      5.2 \\
     1.5 &       79 &      2.6 &     10.5 &        4 &      0.1 \\
     2.0 &       67 &      2.2 &     11.0 &       30 &      1.0 \\
     2.5 &      109 &      3.6 &     .... &      ... &      ... \\
\hline
\end{tabular}
\end{table}

\begin{table}
\centering
\caption{Statistics of Families
}
\label{tab:fstats}
\begin{tabular}{lrrr}
\hline
Family & $F$    &   $n$    & \%$N$     \\
1      &     2  &     3    &   4       \\
\hline
      &         &         &   \\
      & Full Sample & $N$=3416 &  \\
      &         &         &         \\
SA     & 0.0 &   1140 &   33.4$\pm$0.8 \\
\SuAB  & 1.0 &    462 &   13.5$\pm$0.6 \\
SAB    & 2.0 &    658 &   19.3$\pm$0.7 \\
\SAuB  & 3.0 &    422 &   12.4$\pm$0.6 \\
SB     & 4.0 &    734 &   21.5$\pm$0.7 \\
      &         &         &   \\
      & $logR_{25}$ $\leq$ 0.40  & $N$=2440  & \\
      &         &         &         \\
SA     & 0.0 &    812 &   33.3$\pm$1.0 \\
\SuAB  & 1.0 &    321 &   13.2$\pm$0.7 \\
SAB    & 2.0 &    460 &   18.9$\pm$0.8 \\
\SAuB  & 3.0 &    296 &   12.1$\pm$0.7 \\
SB     & 4.0 &    551 &   22.6$\pm$0.8 \\
\hline
\end{tabular}
\end{table}

\begin{table}
\centering
\caption{Bar fraction versus stage
}
\label{tab:barstage}
\begin{tabular}{rcrrcr}
\hline
$T$ & $f_{bar}$(\%)    &   $N(T)$    & $T$ & $f_{bar}$(\%)    &   $N(T)$       \\
1      &     2  &     3    &  1 & 2 & 3       \\
\hline
      &         &         &   \\
      & Full Sample & $N$=3416 &  \\
      &         &         &         \\
 $-$3.0 &   33.3$\pm$5.1 &     84 &   4.0 &   61.8$\pm$   3.6 &    186 \\
 $-$2.5 &   38.3$\pm$7.1 &     47 &   4.5 &   49.8$\pm$   3.1 &    259 \\
 $-$2.0 &   45.5$\pm$5.7 &     77 &   5.0 &   40.6$\pm$   3.1 &    244 \\
 $-$1.5 &   60.7$\pm$6.3 &     61 &   5.5 &   54.1$\pm$   3.7 &    181 \\
 $-$1.0 &   67.0$\pm$4.5 &    109 &   6.0 &   77.0$\pm$   4.2 &    100 \\
 $-$0.5 &   60.3$\pm$6.4 &     58 &   6.5 &   84.5$\pm$   3.0 &    148 \\
  0.0 &   49.4$\pm$5.6 &     79 &   7.0 &   92.4$\pm$   2.3 &    132 \\
  0.5 &   67.0$\pm$4.8 &     97 &   7.5 &   82.3$\pm$   3.6 &    113 \\
  1.0 &   60.0$\pm$4.6 &    115 &   8.0 &   94.1$\pm$   2.2 &    118 \\
  1.5 &   68.1$\pm$4.4 &    113 &   8.5 &   86.4$\pm$   3.7 &     88 \\
  2.0 &   69.4$\pm$4.7 &     98 &   9.0 &   90.5$\pm$   2.6 &    126 \\
  2.5 &   72.4$\pm$3.7 &    145 &   9.5 &   91.2$\pm$   3.4 &     68 \\
  3.0 &   68.5$\pm$3.2 &    216 &  10.0 &   84.9$\pm$   3.3 &    119 \\
  3.5 &   66.8$\pm$3.1 &    235 &  all $T$ &   66.6$\pm$   0.8 &   3416 \\
      &         &         &   \\
      & $logR_{25}$ $\leq$ 0.40  & $N$=2440  & \\
      &         &         &         \\
 $-$3.0 &   33.3$\pm$6.1 &     60 &   4.0 &   61.9$\pm$   4.1 &    139 \\
 $-$2.5 &   37.5$\pm$8.6 &     32 &   4.5 &   47.1$\pm$   3.5 &    208 \\
 $-$2.0 &   46.7$\pm$6.4 &     60 &   5.0 &   41.7$\pm$   3.7 &    175 \\
 $-$1.5 &   64.3$\pm$7.4 &     42 &   5.5 &   50.4$\pm$   4.4 &    131 \\
 $-$1.0 &   69.3$\pm$5.3 &     75 &   6.0 &   79.4$\pm$   4.9 &     68 \\
 $-$0.5 &   71.9$\pm$7.9 &     32 &   6.5 &   88.6$\pm$   3.1 &    105 \\
  0.0 &   51.7$\pm$6.6 &     58 &   7.0 &   91.5$\pm$   3.3 &     71 \\
  0.5 &   59.1$\pm$6.1 &     66 &   7.5 &   75.8$\pm$   5.3 &     66 \\
  1.0 &   65.9$\pm$5.1 &     88 &   8.0 &   93.0$\pm$   2.7 &     86 \\
  1.5 &   65.4$\pm$5.4 &     78 &   8.5 &   83.1$\pm$   4.7 &     65 \\
  2.0 &   73.1$\pm$5.4 &     67 &   9.0 &   90.6$\pm$   3.0 &     96 \\
  2.5 &   73.4$\pm$4.2 &    109 &   9.5 &   88.0$\pm$   4.6 &     50 \\
  3.0 &   73.6$\pm$3.7 &    140 &  10.0 &   83.3$\pm$   3.7 &    102 \\
  3.5 &   70.8$\pm$3.5 &    171 &  all $T$ &   66.7$\pm$   1.0 &   2440 \\
\hline
\end{tabular}
\end{table}

\begin{figure}
\includegraphics[width=\columnwidth]{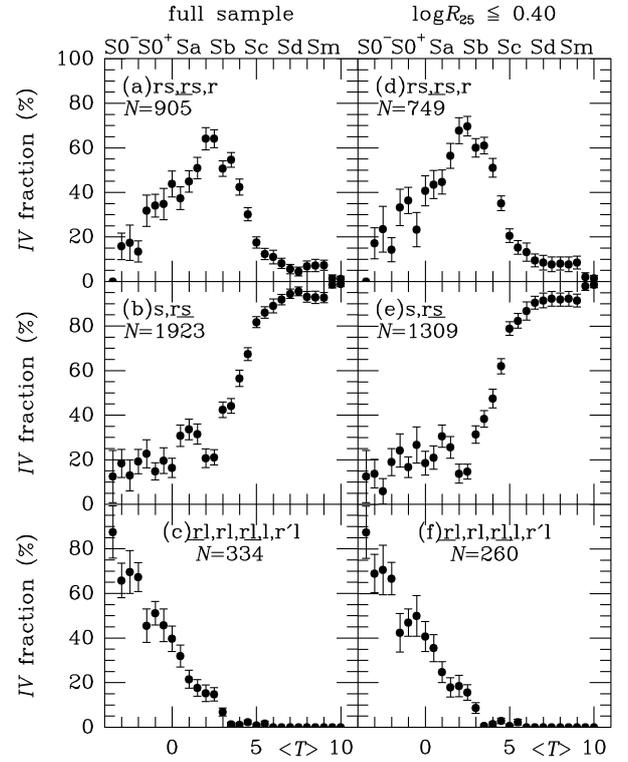}
\vspace{-5truemm}
\caption{Distribution of CVRHS inner varieties versus stage. 
The numbers of objects in each subtype are listed in Table ~\ref{tab:ivstats}}
\label{fig:ringstage}
\end{figure}

\begin{figure}
\includegraphics[width=\columnwidth]{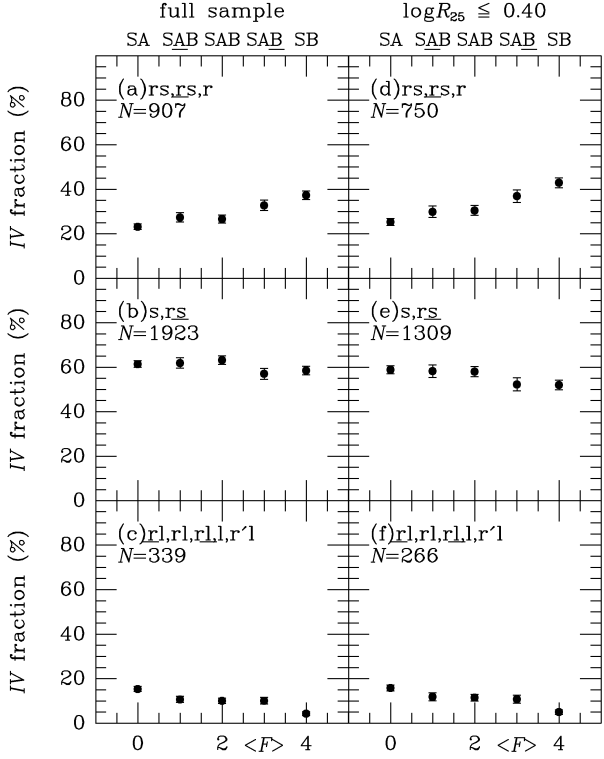}
\vspace{-5truemm}
\caption{Distribution of CVRHS inner varieties versus family
}
\label{fig:ringsbars}
\end{figure}

Figures~\ref{fig:histov}--~\ref{fig:ringsbars} and
Tables~\ref{tab:ivstats}--~\ref{tab:ringfstats1} show the statistics of
CVRHS inner varieties grouped into three categories: inner rings and
pseudorings [(rs), ($\underline{\rm r}$s), and (r)]; pure spirals and
weak inner pseudorings [(s) and (r$\underline{\rm s}$)], and lenses and
ring/pseudoring-lenses [($\underline{\rm r}$l), (rl), r$\underline{\rm
l}$), (l), and (r$^{\prime}$l)]. Figure~\ref{fig:histov} shows that the
most common inner variety is (s), and that inner pseudorings are more
common than closed inner rings and ring-lenses.
Figure~\ref{fig:ringstage} shows the relative frequency of the three
inner variety subgroups versus CVRHS stage, which highlights how the
most prominent inner rings and pseudorings occur near stage Sab (see
also de Lapparent et al. 2011), and that the frequency of such rings
declines rapidly towards earlier and later stages. (s) and
(r$\underline{\rm s}$) varieties become most frequent for stages Sbc
and later, while lenses and ring/pseudoring lenses are most common
among S0s and become infrequent by stage S$\underline{\rm b}$c.
Figure~\ref{fig:ringsbars} shows the distribution of inner varieties by
CVRHS family. These graphs show that although inner rings and
well-defined inner pseudorings are most abundant in SB galaxies and
least abundant in SA galaxies, there is still a significant ring
frequency in SA galaxies. As for the bar fraction statistics, these
trends also have little dependence on whether the sample used is the
full sample or the restricted subset.

\begin{table}
\centering
\caption{Statistics of Inner Varieties
}
\label{tab:ivstats}
\begin{tabular}{lrr}
\hline
Inner Variety &    $n$    & \%$N$        \\
1             &     2     &     3        \\
\hline
      &         &            \\
      & Full Sample & $N$=3162  \\
      &         &                  \\
s   &    1515 &   47.9$\pm$ 0.9 \\
\rus &    408 &   12.9$\pm$ 0.6 \\
rs  &     458 &   14.5$\pm$ 0.6 \\
\urs &    263 &    8.3$\pm$ 0.5 \\
r   &     184 &    5.8$\pm$ 0.4 \\
\url &     50 &    1.6$\pm$ 0.2 \\
rl  &      33 &    1.0$\pm$ 0.2 \\
\rul &     39 &    1.2$\pm$ 0.2 \\
l   &     157 &    5.0$\pm$ 0.4 \\
\irpl &    55 &    1.7$\pm$ 0.2 \\
      &         &            \\
      & $logR_{25}$ $\leq$ 0.40  & $N$=2318  \\
      &         &                  \\
s   &     982 &   42.4$\pm$ 1.0 \\
\rus &    327 &   14.1$\pm$ 0.7 \\
rs  &     391 &   16.9$\pm$ 0.8 \\
\urs &    224 &    9.7$\pm$ 0.6 \\
r   &     134 &    5.8$\pm$ 0.5 \\
\url &     43 &    1.9$\pm$ 0.3 \\
rl  &      19 &    0.8$\pm$ 0.2 \\
\rul &     32 &    1.4$\pm$ 0.2 \\
l   &     119 &    5.1$\pm$ 0.5 \\
\irpl &    47 &    2.0$\pm$ 0.3 \\
\hline
\end{tabular}
\end{table}

\begin{table}
\centering
\caption{Statistics of Inner Varieties by stage - full sample
}
\label{tab:ringtstats1}
\begin{tabular}{rrrrr}
\hline
$T$ &    $n$    & rs,\urs,r & s,\rus & \url,rl,\rul,l,\irpl \\
    &           & $f$\% & $f$\% & $f$\% \\
1             &     2     &     3    & 4 & 5    \\
\hline
  $-$3.5 &      8 &    0.0$\pm$ 0.0 &   12.5$\pm$11.7 &   87.5$\pm$11.7 \\
  $-$3.0 &     38 &   15.8$\pm$ 5.9 &   18.4$\pm$ 6.3 &   65.8$\pm$ 7.7 \\
  $-$2.5 &     23 &   17.4$\pm$ 7.9 &   13.0$\pm$ 7.0 &   69.6$\pm$ 9.6 \\
  $-$2.0 &     52 &   13.5$\pm$ 4.7 &   19.2$\pm$ 5.5 &   67.3$\pm$ 6.5 \\
  $-$1.5 &     44 &   31.8$\pm$ 7.0 &   22.7$\pm$ 6.3 &   45.5$\pm$ 7.5 \\
  $-$1.0 &     88 &   34.1$\pm$ 5.1 &   14.8$\pm$ 3.8 &   51.1$\pm$ 5.3 \\
  $-$0.5 &     46 &   34.8$\pm$ 7.0 &   19.6$\pm$ 5.8 &   45.7$\pm$ 7.3 \\
   0.0 &     73 &   43.8$\pm$ 5.8 &   16.4$\pm$ 4.3 &   39.7$\pm$ 5.7 \\
   0.5 &     91 &   37.4$\pm$ 5.1 &   30.8$\pm$ 4.8 &   31.9$\pm$ 4.9 \\
   1.0 &    107 &   44.9$\pm$ 4.8 &   33.6$\pm$ 4.6 &   21.5$\pm$ 4.0 \\
   1.5 &    108 &   50.9$\pm$ 4.8 &   31.5$\pm$ 4.5 &   17.6$\pm$ 3.7 \\
   2.0 &     92 &   64.1$\pm$ 5.0 &   20.7$\pm$ 4.2 &   15.2$\pm$ 3.7 \\
   2.5 &    142 &   64.1$\pm$ 4.0 &   21.1$\pm$ 3.4 &   14.8$\pm$ 3.0 \\
   3.0 &    205 &   50.7$\pm$ 3.5 &   42.4$\pm$ 3.5 &    6.8$\pm$ 1.8 \\
   3.5 &    229 &   54.6$\pm$ 3.3 &   44.1$\pm$ 3.3 &    1.3$\pm$ 0.8 \\
   4.0 &    184 &   42.4$\pm$ 3.6 &   56.5$\pm$ 3.7 &    1.1$\pm$ 0.8 \\
   4.5 &    258 &   30.2$\pm$ 2.9 &   67.4$\pm$ 2.9 &    2.3$\pm$ 0.9 \\
   5.0 &    240 &   17.5$\pm$ 2.5 &   81.7$\pm$ 2.5 &    0.8$\pm$ 0.6 \\
   5.5 &    178 &   12.4$\pm$ 2.5 &   86.0$\pm$ 2.6 &    1.7$\pm$ 1.0 \\
   6.0 &    100 &   11.0$\pm$ 3.1 &   89.0$\pm$ 3.1 &    0.0$\pm$ 0.0 \\
   6.5 &    146 &    8.2$\pm$ 2.3 &   91.8$\pm$ 2.3 &    0.0$\pm$ 0.0 \\
   7.0 &    125 &    5.6$\pm$ 2.1 &   94.4$\pm$ 2.1 &    0.0$\pm$ 0.0 \\
   7.5 &    111 &    4.5$\pm$ 2.0 &   95.5$\pm$ 2.0 &    0.0$\pm$ 0.0 \\
   8.0 &    116 &    6.9$\pm$ 2.4 &   93.1$\pm$ 2.4 &    0.0$\pm$ 0.0 \\
   8.5 &     83 &    7.2$\pm$ 2.8 &   92.8$\pm$ 2.8 &    0.0$\pm$ 0.0 \\
   9.0 &    123 &    7.3$\pm$ 2.3 &   92.7$\pm$ 2.3 &    0.0$\pm$ 0.0 \\
   9.5 &     67 &    1.5$\pm$ 1.5 &   98.5$\pm$ 1.5 &    0.0$\pm$ 0.0 \\
  10.0 &     85 &    1.2$\pm$ 1.2 &   98.8$\pm$ 1.2 &    0.0$\pm$ 0.0 \\
all $T$ &   3162 &   28.6$\pm$ 0.8 &   60.8$\pm$ 0.9 &   10.6$\pm$ 0.5 \\
\hline
\end{tabular}
\end{table}

\begin{table}
\centering
\caption{Statistics of Inner Varieties by stage - restricted sample
}
\label{tab:ringtstats2}
\begin{tabular}{rrrrr}
\hline
$T$ &    $n$    & rs,\urs,r & s,\rus & \url,rl,\rul,l,\irpl \\
    &           & $f$\% & $f$\% & $f$\% \\
1             &     2     &     3    & 4 & 5    \\
\hline
  $-$3.5 &      8 &    0.0$\pm$ 0.0 &   12.5$\pm$11.7 &   87.5$\pm$11.7 \\
  $-$3.0 &     29 &   17.2$\pm$ 7.0 &   13.8$\pm$ 6.4 &   69.0$\pm$ 8.6 \\
  $-$2.5 &     17 &   23.5$\pm$10.3 &    5.9$\pm$ 5.7 &   70.6$\pm$11.1 \\
  $-$2.0 &     42 &   14.3$\pm$ 5.4 &   19.0$\pm$ 6.1 &   66.7$\pm$ 7.3 \\
  $-$1.5 &     33 &   33.3$\pm$ 8.2 &   24.2$\pm$ 7.5 &   42.4$\pm$ 8.6 \\
  $-$1.0 &     66 &   36.4$\pm$ 5.9 &   16.7$\pm$ 4.6 &   47.0$\pm$ 6.1 \\
  $-$0.5 &     30 &   23.3$\pm$ 7.7 &   26.7$\pm$ 8.1 &   50.0$\pm$ 9.1 \\
   0.0 &     54 &   40.7$\pm$ 6.7 &   18.5$\pm$ 5.3 &   40.7$\pm$ 6.7 \\
   0.5 &     62 &   43.5$\pm$ 6.3 &   21.0$\pm$ 5.2 &   35.5$\pm$ 6.1 \\
   1.0 &     85 &   44.7$\pm$ 5.4 &   30.6$\pm$ 5.0 &   24.7$\pm$ 4.7 \\
   1.5 &     78 &   56.4$\pm$ 5.6 &   25.6$\pm$ 4.9 &   17.9$\pm$ 4.3 \\
   2.0 &     65 &   67.7$\pm$ 5.8 &   13.8$\pm$ 4.3 &   18.5$\pm$ 4.8 \\
   2.5 &    109 &   69.7$\pm$ 4.4 &   14.7$\pm$ 3.4 &   15.6$\pm$ 3.5 \\
   3.0 &    140 &   60.0$\pm$ 4.1 &   31.4$\pm$ 3.9 &    8.6$\pm$ 2.4 \\
   3.5 &    172 &   61.0$\pm$ 3.7 &   38.4$\pm$ 3.7 &    0.6$\pm$ 0.6 \\
   4.0 &    139 &   51.1$\pm$ 4.2 &   47.5$\pm$ 4.2 &    1.4$\pm$ 1.0 \\
   4.5 &    208 &   35.1$\pm$ 3.3 &   62.0$\pm$ 3.4 &    2.9$\pm$ 1.2 \\
   5.0 &    175 &   20.6$\pm$ 3.1 &   78.9$\pm$ 3.1 &    0.6$\pm$ 0.6 \\
   5.5 &    131 &   15.3$\pm$ 3.1 &   82.4$\pm$ 3.3 &    2.3$\pm$ 1.3 \\
   6.0 &     68 &   13.2$\pm$ 4.1 &   86.8$\pm$ 4.1 &    0.0$\pm$ 0.0 \\
   6.5 &    105 &    9.5$\pm$ 2.9 &   90.5$\pm$ 2.9 &    0.0$\pm$ 0.0 \\
   7.0 &     71 &    8.5$\pm$ 3.3 &   91.5$\pm$ 3.3 &    0.0$\pm$ 0.0 \\
   7.5 &     65 &    7.7$\pm$ 3.3 &   92.3$\pm$ 3.3 &    0.0$\pm$ 0.0 \\
   8.0 &     86 &    8.1$\pm$ 2.9 &   91.9$\pm$ 2.9 &    0.0$\pm$ 0.0 \\
   8.5 &     65 &    7.7$\pm$ 3.3 &   92.3$\pm$ 3.3 &    0.0$\pm$ 0.0 \\
   9.0 &     94 &    8.5$\pm$ 2.9 &   91.5$\pm$ 2.9 &    0.0$\pm$ 0.0 \\
   9.5 &     50 &    2.0$\pm$ 2.0 &   98.0$\pm$ 2.0 &    0.0$\pm$ 0.0 \\
  10.0 &     71 &    1.4$\pm$ 1.4 &   98.6$\pm$ 1.4 &    0.0$\pm$ 0.0 \\
all $T$ &   2318 &   32.3$\pm$ 1.0 &   56.5$\pm$ 1.0 &   11.2$\pm$ 0.7 \\
\hline
\end{tabular}
\end{table}

\begin{table}
\centering
\caption{Statistics of Inner Varieties by family
}
\label{tab:ringfstats1}
\begin{tabular}{rrrrr}
\hline
$F$ &    $n$    & rs,\urs,r & s,\rus & \url,rl,\rul,l,\irpl \\
    &           & $f$\% & $f$\% & $f$\% \\
1             &     2     &     3    & 4 & 5    \\
\hline
      &         &            \\
      & Full Sample & $N$=3169  \\
      &         &                  \\
   0.0 &   1021 &   23.2$\pm$ 1.3 &   61.4$\pm$ 1.5 &   15.4$\pm$ 1.1 \\
   1.0 &    457 &   27.4$\pm$ 2.1 &   61.9$\pm$ 2.3 &   10.7$\pm$ 1.4 \\
   2.0 &    631 &   26.6$\pm$ 1.8 &   63.2$\pm$ 1.9 &   10.1$\pm$ 1.2 \\
   3.0 &    405 &   32.8$\pm$ 2.3 &   57.0$\pm$ 2.5 &   10.1$\pm$ 1.5 \\
   4.0 &    655 &   37.3$\pm$ 1.9 &   58.5$\pm$ 1.9 &    4.3$\pm$ 0.8 \\
all $F$ &   3169 &   28.6$\pm$ 0.8 &   60.7$\pm$ 0.9 &   10.7$\pm$ 0.5 \\
      &         &            \\
      & $logR_{25}$ $\leq$ 0.40  & $N$=2325  \\
      &         &                  \\
   0.0 &    760 &   25.3$\pm$ 1.6 &   58.9$\pm$ 1.8 &   15.8$\pm$ 1.3 \\
   1.0 &    318 &   29.9$\pm$ 2.6 &   58.2$\pm$ 2.8 &   11.9$\pm$ 1.8 \\
   2.0 &    443 &   30.5$\pm$ 2.2 &   58.0$\pm$ 2.3 &   11.5$\pm$ 1.5 \\
   3.0 &    287 &   36.9$\pm$ 2.8 &   52.3$\pm$ 2.9 &   10.8$\pm$ 1.8 \\
   4.0 &    517 &   42.9$\pm$ 2.2 &   52.0$\pm$ 2.2 &    5.0$\pm$ 1.0 \\
all $F$ &   2325 &   32.3$\pm$ 1.0 &   56.3$\pm$ 1.0 &   11.4$\pm$ 0.7 \\
\hline
\end{tabular}
\end{table}

\begin{figure}
\includegraphics[width=\columnwidth]{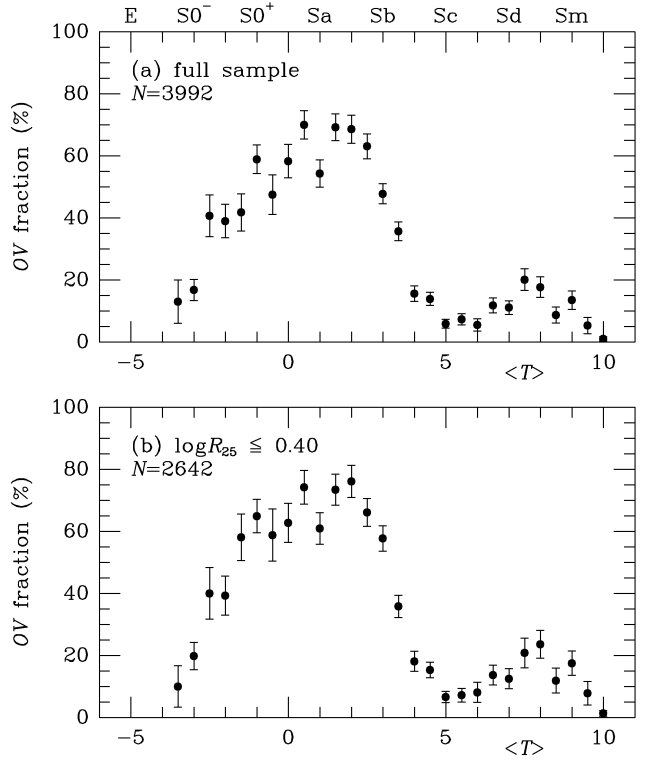}
\vspace{-5truemm}
\caption{Outer variety fraction as a function of CVRHS stage
}
\label{fig:OV-all-types}
\end{figure}

Table~\ref{tab:OVstats} summarizes the numbers of different kinds of
outer features recognized in CVRHS classifications of EFIGI galaxies,
including what Buta (2017b) refers to as ``outer resonant subclasses":
R$_1$, R$_1^{\prime}$, R$_1$R$_2^{\prime}$, and R$_2^{\prime}$, of
which 172 examples are included. Other types of outer features include
outer lenses (L), outer rings (R), and outer ring lenses (RL) (see
section 8.9), outer pseudorings (R$^{\prime}$) not clearly in any other
subcategory, outer pseudoring-lenses (R$^{\prime}$L), and a variety of
multiple outer feature categories such as a doubled outer ring (RR) and
doubled outer pseudorings (R$^{\prime}$,R$^{\prime}$). Only 24\% of the
galaxies in the full EFIGI sample are recorded as having an outer
feature in Table~\ref{tab:catalog}. Figure~\ref{fig:OV-all-types} shows
the percentages of galaxies having an outer feature as a function of
CVRHS stage. Like inner rings and pseudorings, outer features are most
common near stage Sab and decrease in frequency towards earlier or
later types, having the lowest frequency in Sc galaxies (see also de
Lapparent et al. 2011). Surprisingly, the relative frequency of outer
features has a secondary peak at stages later than Sc. Consistent with
Comer\'on et al. (2014), outer features in the EFIGI sample are
infrequent at stages later than Sb.

Table~\ref{tab:OV-meanT} highlights the mean stage associated with 10
outer feature types or groups of types having 10 or more galaxies.
Outer lenses (L) are the main outer features characteristic of
early-type disk galaxies (S0$^o$ to S0$^+$), while outer pseudorings
are characteristic of much later types (S$\underline{\rm a}$b to Scd).
The outer resonant subclasses are found in the range S0/a to
Sa$\underline{\rm b}$, with \RoneP\ cases averaging about half a stage
earlier than \RtwoP\ cases. This range is similar to that found by de
Lapparent et al. (2011). Outer ring-lenses (RL) average at stage S0/a
(S0$^-$ to Sb), while outer rings (R) average at stage Sa (S0$^o$ to
S$\underline{\rm b}$c).  Table~\ref{tab:OV-meanT} also summarizes the
mean family classification for each outer feature type. The weakest
bars are found for outer rings, lenses, ring-lenses, and pseudorings,
while the strongest bars are found for the outer resonant subclasses,
especially for type R$_1^{\prime}$.

\begin{figure}
\includegraphics[width=\columnwidth]{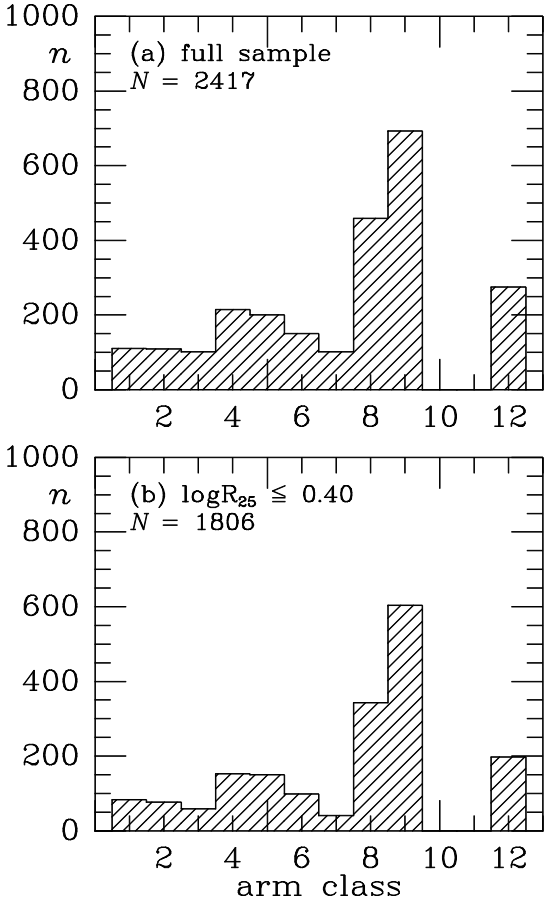} 
\vspace{-5truemm}
\caption{Distribution of Elmegreen Arm Classes } 
\label{fig:armclasses}
\end{figure}

\begin{figure}
\includegraphics[width=\columnwidth]{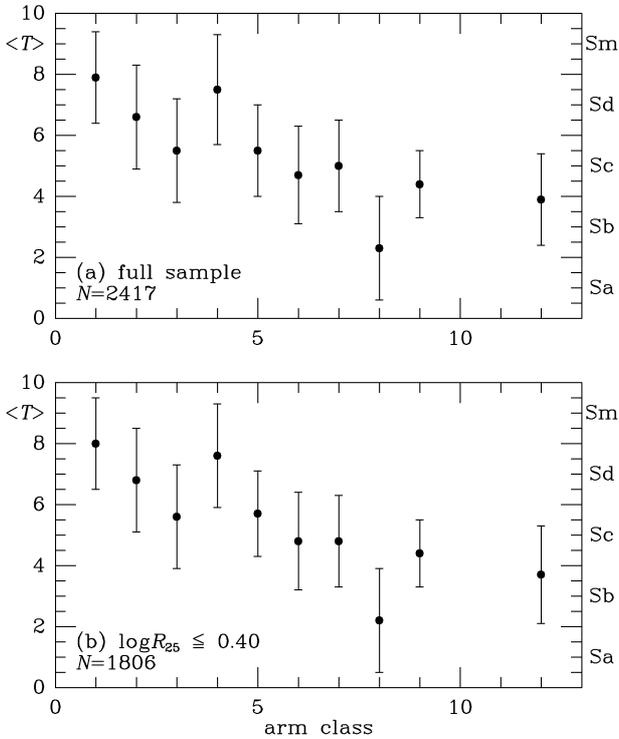}
\vspace{-5truemm}
\caption{Mean stages of Elmegreen Arm Classes
}
\label{fig:stageAC}
\end{figure}

The distribution of arm classes, as defined by Elmegreen and Elmegreen
(1987), is shown in Figure~\ref{fig:armclasses} and
Table~\ref{tab:acstats}. AC 1--4 include mainly flocculent spirals,
while AC 5--12  include mainly grand design spirals. The sample
includes 21--22\% of the flocculent types and 78--79\% of the grand
design types. Of the latter, most are grand design multi-armed spirals
which account for 29\% of the full sample and 33\% of the restricted
sample. Figure~\ref{fig:stageAC} shows the mean stage as a function of
arm class.  The flocculent categories average between stages
S$\underline{\rm c}$d to Sdm, while the strongest grand design
categories average between stages Sab and Sbc. This implies that AC1-4
galaxies are generally less luminous than AC 8-12 galaxies, as shown
by Figure 2 of de Lapparent et al. (2011).

\begin{figure}
\includegraphics[width=\columnwidth]{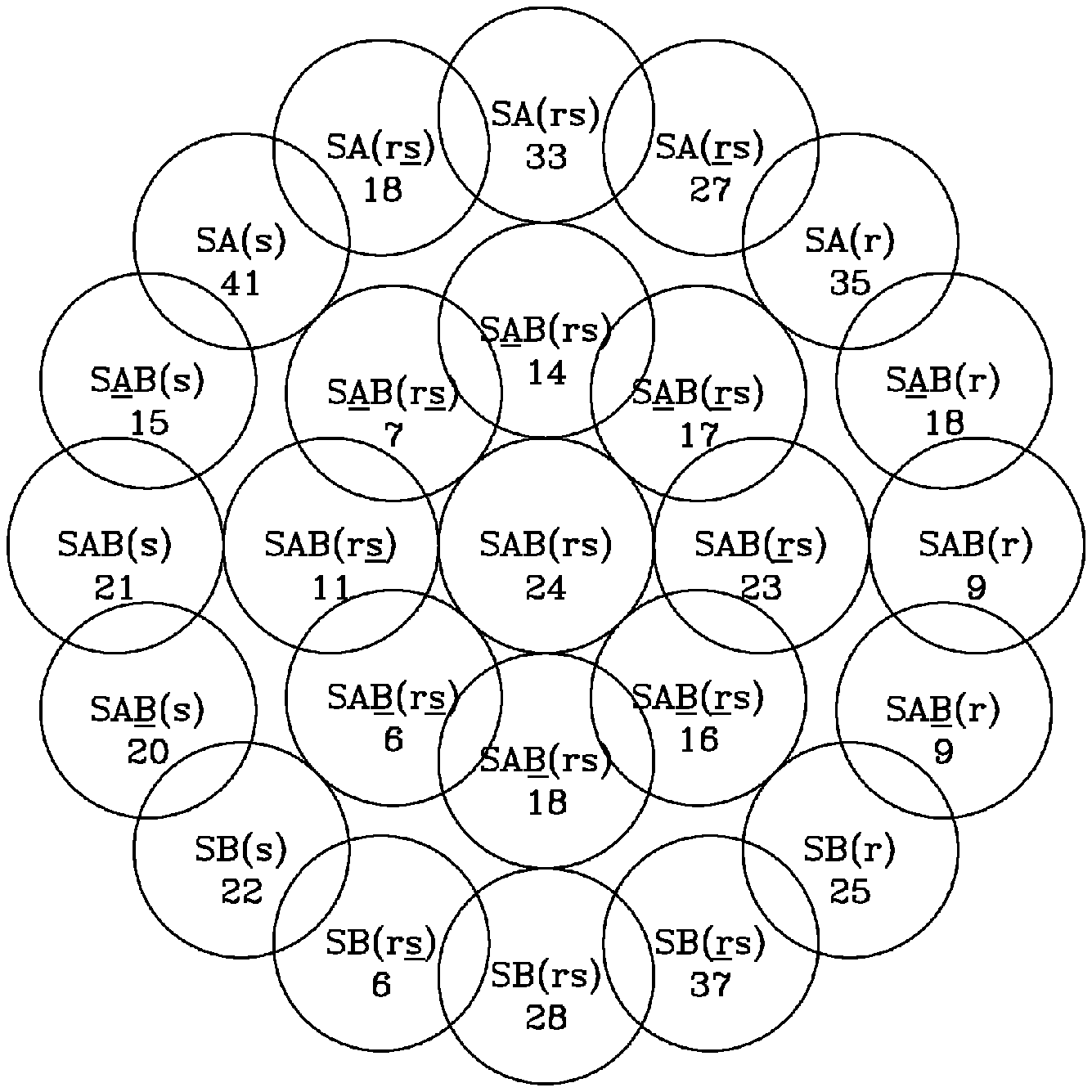} 
\vspace{-5truemm}
\caption{Distribution of EFIGI galaxies in cells of the CVRHS system from
stages S$\underline{\rm 0}$/a to Sa$\underline{\rm b}$.} 
\label{fig:stage1}
\end{figure}

\begin{figure}
\includegraphics[width=\columnwidth]{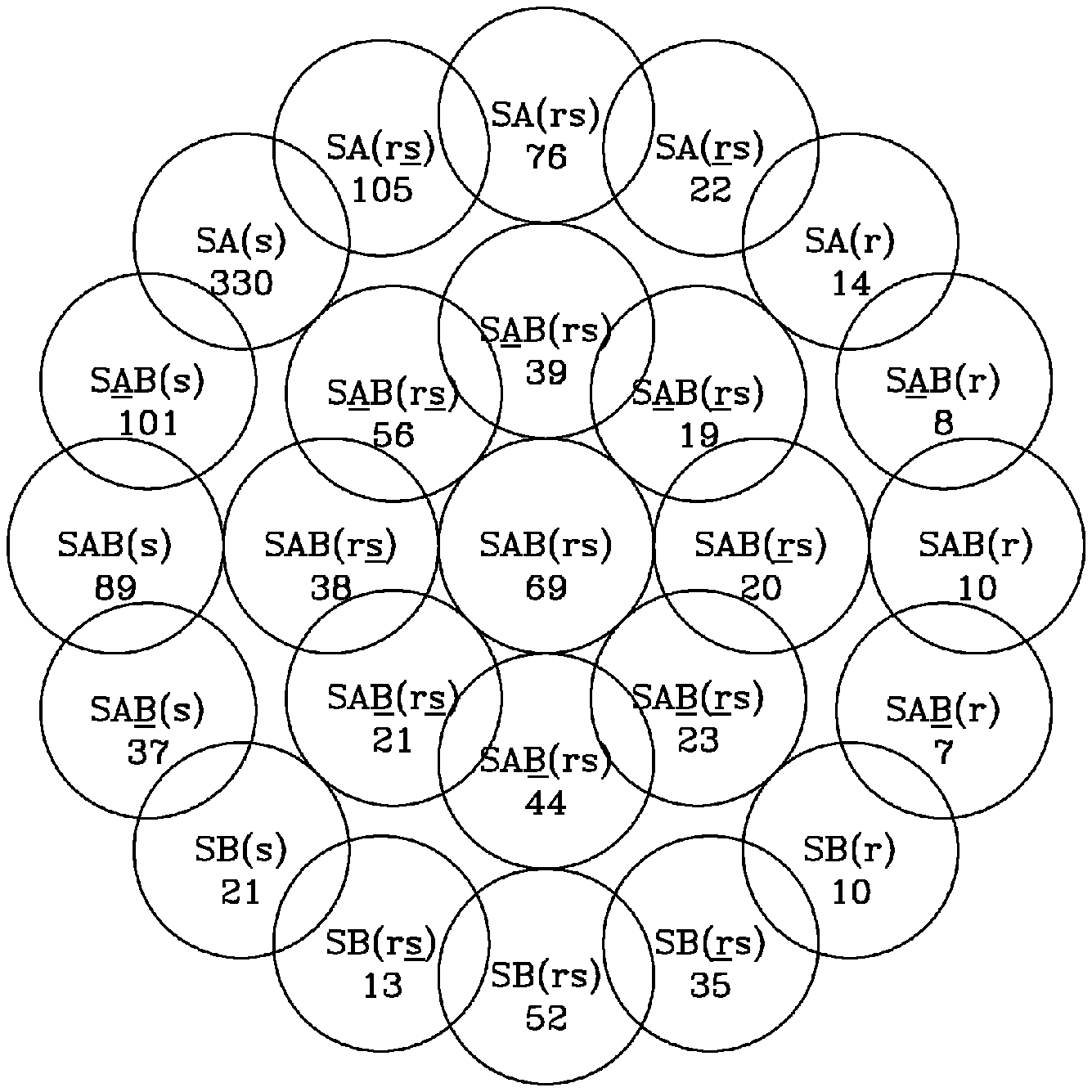} 
\vspace{-5truemm}
\caption{Distribution of EFIGI galaxies in cells of the CVRHS system from
stages Sb to S$\underline{\rm c}$d.} 
\label{fig:stage2}
\end{figure}

\begin{figure}
\includegraphics[width=\columnwidth]{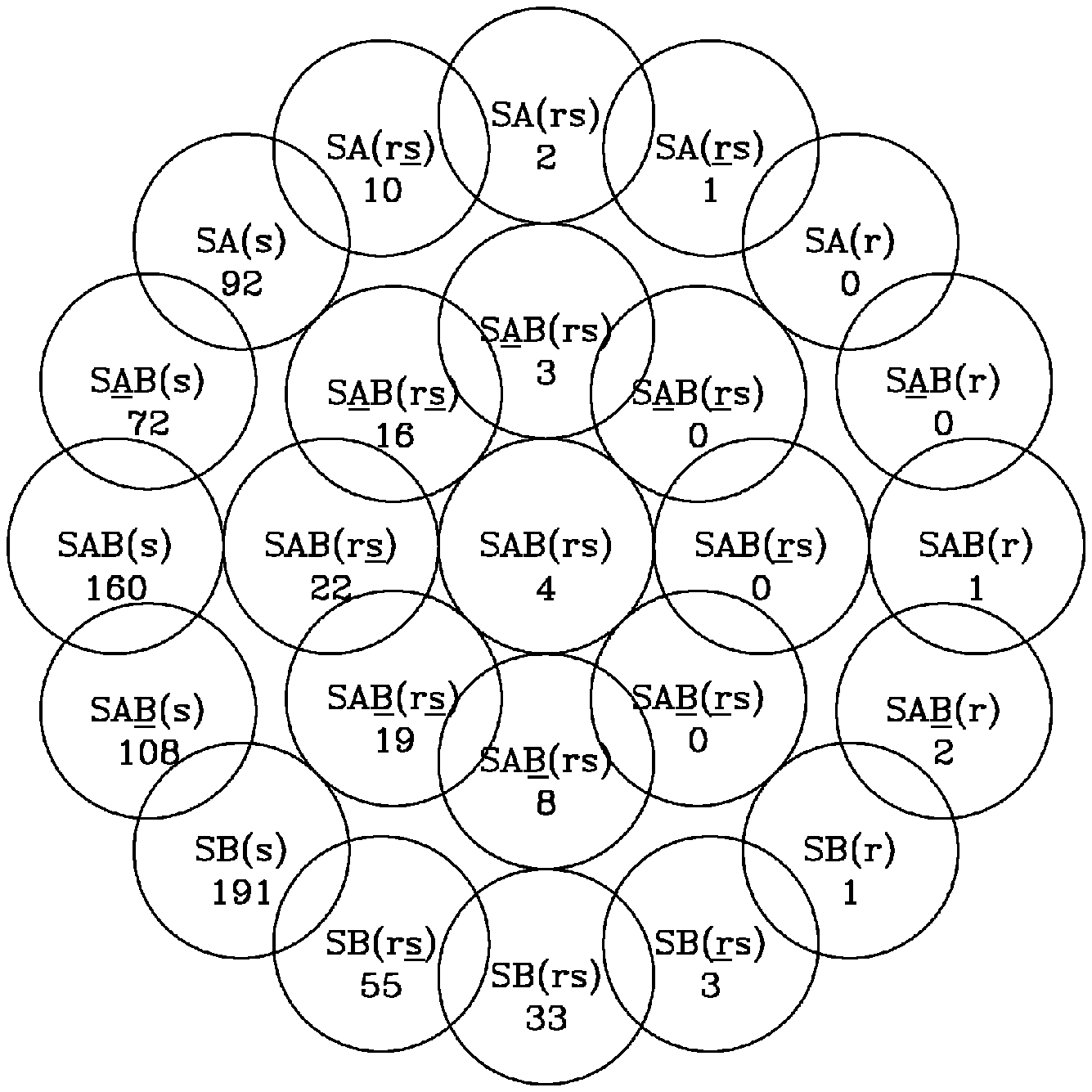} 
\vspace{-5truemm}
\caption{Distribution of EFIGI galaxies in cells of the CVRHS system from
stages Scd to Sm.} 
\label{fig:stage3}
\end{figure}

An important issue that can be examined with the EFIGI sample is the
relative populations of the different ``cells" of the CVRHS
classification volume. A cell generally consists of a combination of
stage, family, and variety, examples being SB(s)cd and S$\underline{\rm
A}$B(r$\underline{\rm s}$)b. For this analysis, additional features such
as outer varieties, lenses, nuclear structures, ansae bars, X patterns,
and extra inner varieties are ignored. This leaves 25 cells at each
CVRHS stage. The relative populations of these cells for the EFIGI sample
are shown in Figures~\ref{fig:stage1}-~\ref{fig:stage3} for three CVRHS stage
intervals: S$\underline{\rm 0}$/a to Sa$\underline{\rm b}$ (early
subgroup), Sb to S$\underline{\rm c}$d (intermediate subgroup), and Scd
to Sm (late subgroup). From these it is clear that the cells of the
CVRHS system are not uniformly populated. There are distinct
nonuniformities: the intermediate subgroup emphasizes cell SA(s) while
the late subgroup emphasizes cell SB(s). Only for the
early subgroup are the populations of the cells relatively uniform.

\begin{figure}
\includegraphics[width=\columnwidth]{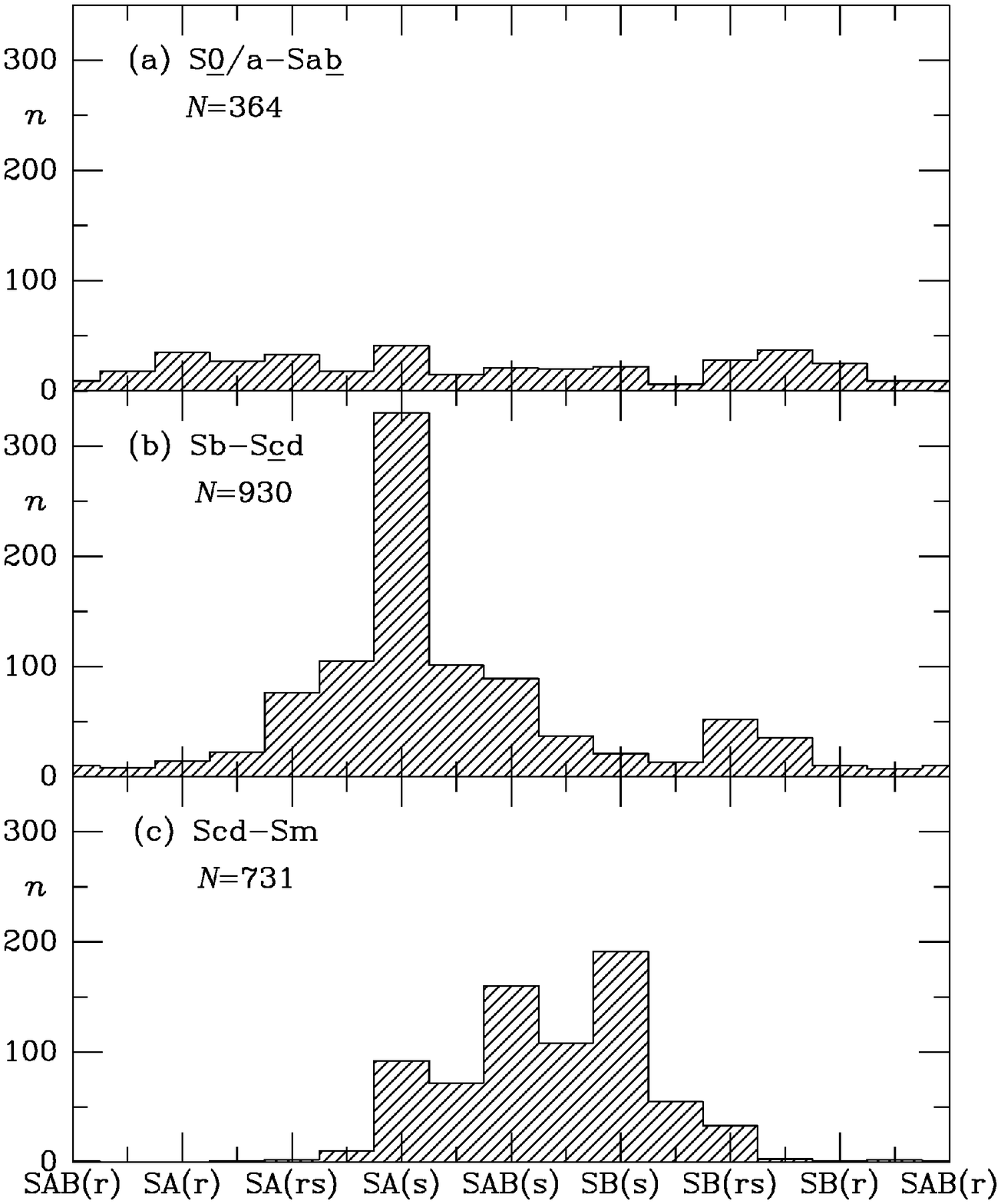} 
\vspace{-5truemm}
\caption{Histograms of the number of EFIGI galaxies within the 16 outer
cross-sectional cells of the CVRHS.} 
\label{fig:outer-circles}
\end{figure}

The histograms in Figure~\ref{fig:outer-circles} show these trends
another way by plotting the number of galaxies in the 16 outer cells of
each stage subgroup going counter-clockwise from cell SAB(r). These
show that the diversity of spiral galaxy morphology diminishes with
advancing stage. The original de Vaucouleurs (1959) classification
volume tried to show this by narrowing down the volume from stage S0/a
to stage Im.

\begin{table}
\centering
\caption{Statistics of Outer Features
}
\label{tab:OVstats}
\begin{tabular}{lrlrlr}
\hline
Feature         &   $n$ & Feature         &   $n$ & Feature              &   $n$ \\
1               &     2 & 1               &     2 & 1                    &     2 \\
\hline
(L)             &   116 & (\RoneP)        &    95 & (\Rprime,R)          &     7 \\
(L:)            &     1 & (\RonePL)       &     4 & (\Rprime,\RoneRtwoP) &     1 \\
(L,R)           &     1 & (\RonePRtwoP)   &     4 & (\Rprime,RL)         &     1 \\
(L,RL)          &     1 & (\RoneRtwo)     &     1 & (\Rprime,\Rprime)    &    13 \\
(L,\Rprime)     &     3 & (\RoneRtwoP)    &    19 & (\Rprime :,\RtwoP)   &     1 \\
(PR)            &     1 & (\RtwoP)        &    37 & (\RprimeL)           &    73 \\
(R)             &    65 & (RL)            &    58 & (\RprimeL,RL)        &     1 \\
(R:)            &     4 & (RL,L)          &     1 & (\RprimeL,\Rprime)   &     1 \\
(R?)            &     1 & (RL,R)          &     1 & (\RprimeL,\RprimeL)  &     1 \\
(R,\Rprime)     &     2 & (RL,\Rprime)    &     1 & (RR)                 &     2 \\
(\Rone)         &     9 & (RL?)           &     1 & \cRuL                &     4 \\
(\RonetwoP)     &     1 & (\Rprime)       &   551 & (R$_{dust}$)         &     1 \\
(\RoneL)        &     2 & (\Rprime,L)     &     1 & no outer feature     &  3371 \\
\hline
\end{tabular}
\end{table}

\begin{table}
\centering
\caption{Mean stages and families for outer features of different types
}
\label{tab:OV-meanT}
\begin{tabular}{lrrr}
\hline
Feature         &   $<T>$ & $\sigma_1$  &   $n$  \\
1                    &      2 &      3 &      4  \\
\hline
(L)                  &   -1.3 &    1.9 &    115 \\
(\Rone ), (\RoneL )  &   -0.2 &    1.2 &     11 \\
(RL)                 &    0.0 &    2.8 &     58 \\
(R)                  &    0.7 &    2.8 &     65 \\
(\RoneRtwoP ),(\RonePRtwoP ),(\RoneRtwo )        &    1.5 &    1.3 &     24 \\
(\RoneP )            &    2.0 &    1.1 &     95 \\
(\RprimeL )          &    2.0 &    2.3 &     73 \\
(\RtwoP )            &    2.6 &    1.5 &     37 \\
(\Rprime,\Rprime )   &    3.0 &    1.4 &     13 \\
(\Rprime )           &    3.7 &    2.4 &    550 \\
\hline
Feature         &   $<F>$ & $\sigma_1$  &   $n$  \\
1                    &      2 &      3 &      4  \\
\hline
(\RprimeL )          &    1.3 &    1.4 &     73 \\
(\Rprime,\Rprime )   &    1.5 &    1.5 &     13 \\
(RL)                 &    1.7 &    1.5 &     55 \\
(\Rprime )           &    1.8 &    1.5 &    547 \\
(R)                  &    1.8 &    1.4 &     65 \\
(L)                  &    1.9 &    1.7 &    115 \\
(\RtwoP )            &    2.2 &    1.0 &     37 \\
(\RoneRtwoP ),(\RonePRtwoP ),(\RoneRtwo ) &    2.6 &    1.3 &     24 \\
(\Rone ), (\RoneL )  &    2.6 &    1.2 &     11 \\
(\RoneP )            &    3.0 &    0.9 &     95 \\
\hline
\end{tabular}
\end{table}

\begin{table}
\centering
\caption{Outer Variety fraction versus stage
}
\label{tab:OVstage}
\begin{tabular}{rcrrcr}
\hline
$T$ & $f_{bar}$(\%)    &   $N(T)$    & $T$ & $f_{bar}$(\%)    &   $N(T)$       \\
1      &     2  &     3    &  1 & 2 & 3       \\
\hline
      &         &         &   \\
      & Full Sample & $N$=3992 &  \\
      &         &         &         \\
 -3.5 &   13.0$\pm$7.0 &     23 &   4.0 &   15.6$\pm$   2.5 &    211 \\
 -3.0 &   16.8$\pm$3.4 &    119 &   4.5 &   13.9$\pm$   2.1 &    280 \\
 -2.5 &   40.7$\pm$6.7 &     54 &   5.0 &    5.9$\pm$   1.4 &    289 \\
 -2.0 &   39.0$\pm$5.4 &     82 &   5.5 &    7.3$\pm$   1.8 &    205 \\
 -1.5 &   41.8$\pm$6.0 &     67 &   6.0 &    5.5$\pm$   2.0 &    128 \\
 -1.0 &   58.9$\pm$4.6 &    112 &   6.5 &   11.8$\pm$   2.4 &    178 \\
 -0.5 &   47.5$\pm$6.4 &     61 &   7.0 &   11.1$\pm$   2.2 &    207 \\
  0.0 &   58.3$\pm$5.4 &     84 &   7.5 &   20.1$\pm$   3.5 &    134 \\
  0.5 &   70.0$\pm$4.6 &    100 &   8.0 &   17.7$\pm$   3.3 &    130 \\
  1.0 &   54.3$\pm$4.4 &    129 &   8.5 &    8.7$\pm$   2.6 &    115 \\
  1.5 &   69.2$\pm$4.3 &    117 &   9.0 &   13.5$\pm$   3.0 &    133 \\
  2.0 &   67.9$\pm$4.5 &    106 &   9.5 &    5.3$\pm$   2.6 &     76 \\
  2.5 &   63.1$\pm$4.0 &    149 &  10.0 &    1.0$\pm$   0.7 &    203 \\
  3.0 &   48.0$\pm$3.2 &    248 & all $T$ &   27.2$\pm$   0.7 &   3992 \\
  3.5 &   35.7$\pm$3.0 &    252 &  .... &   ..............  &   .... \\
      &         &         &   \\
      & $logR_{25}$ $\leq$ 0.40  & $N$=2642  & \\
      &         &         &         \\
 -3.5 &   10.0$\pm$6.7 &     20 &   4.0 &   18.1$\pm$   3.2 &    144 \\
 -3.0 &   19.8$\pm$4.4 &     81 &   4.5 &   15.3$\pm$   2.5 &    209 \\
 -2.5 &   40.0$\pm$8.3 &     35 &   5.0 &    6.6$\pm$   1.8 &    183 \\
 -2.0 &   39.3$\pm$6.3 &     61 &   5.5 &    7.2$\pm$   2.2 &    139 \\
 -1.5 &   58.1$\pm$7.5 &     43 &   6.0 &    8.1$\pm$   3.2 &     74 \\
 -1.0 &   64.9$\pm$5.4 &     77 &   6.5 &   13.7$\pm$   3.2 &    117 \\
 -0.5 &   58.8$\pm$8.4 &     34 &   7.0 &   12.5$\pm$   3.2 &    104 \\
  0.0 &   62.7$\pm$6.3 &     59 &   7.5 &   20.8$\pm$   4.8 &     72 \\
  0.5 &   74.2$\pm$5.4 &     66 &   8.0 &   23.6$\pm$   4.5 &     89 \\
  1.0 &   60.9$\pm$5.1 &     92 &   8.5 &   11.9$\pm$   4.0 &     67 \\
  1.5 &   73.4$\pm$5.0 &     79 &   9.0 &   17.5$\pm$   3.9 &     97 \\
  2.0 &   76.1$\pm$5.2 &     67 &   9.5 &    7.8$\pm$   3.8 &     51 \\
  2.5 &   66.1$\pm$4.5 &    109 &  10.0 &    1.3$\pm$   0.9 &    158 \\
  3.0 &   57.7$\pm$4.1 &    142 & all $T$ &   30.3$\pm$   0.9 &   2642 \\
  3.5 &   35.8$\pm$3.6 &    173 &   .... &  ............... &   .... \\
\hline
\end{tabular}
\end{table}

\begin{table}
\centering
\caption{Statistics of Arm Classes
}
\label{tab:acstats}
\begin{tabular}{lrrrr}
\hline
Arm Class &    $n$    & \%$N$  & $<T>$ & $\sigma_1$      \\
1             &     2     &     3     & 4 & 5     \\
\hline
                     &        &                &        &      \\
                     &        & full sample ($N$=2417) &  &   \\
                     &        &                &        &      \\
AC  1                &    110 &    4.6$\pm$0.4 &    7.9 &  1.5 \\
AC  2                &    109 &    4.5$\pm$0.4 &    6.6 &  1.7 \\
AC  3                &    102 &    4.2$\pm$0.4 &    5.5 &  1.7 \\
AC  4                &    215 &    8.9$\pm$0.6 &    7.5 &  1.8 \\
AC  5                &    201 &    8.3$\pm$0.6 &    5.5 &  1.5 \\
AC  6                &    151 &    6.2$\pm$0.5 &    4.7 &  1.6 \\
AC  7                &    102 &    4.2$\pm$0.4 &    5.0 &  1.5 \\
AC  8                &    459 &   19.0$\pm$0.8 &    2.3 &  1.7 \\
AC  9                &    693 &   28.7$\pm$0.9 &    4.4 &  1.1 \\
AC 12                &    275 &   11.4$\pm$0.6 &    3.9 &  1.5 \\
                     &        &                &        &      \\
                     &        & $logR_{25}$ $\leq$ 0.40 ($N$=1806) &  &   \\
                     &        &                &        &      \\
AC  1                &     83 &    4.6$\pm$0.5 &    8.0 &  1.5 \\
AC  2                &     77 &    4.3$\pm$0.5 &    6.8 &  1.7 \\
AC  3                &     59 &    3.3$\pm$0.4 &    5.6 &  1.7 \\
AC  4                &    153 &    8.5$\pm$0.7 &    7.6 &  1.7 \\
AC  5                &    150 &    8.3$\pm$0.6 &    5.7 &  1.4 \\
AC  6                &     99 &    5.5$\pm$0.5 &    4.8 &  1.6 \\
AC  7                &     41 &    2.3$\pm$0.4 &    4.8 &  1.5 \\
AC  8                &    343 &   19.0$\pm$0.9 &    2.2 &  1.7 \\
AC  9                &    604 &   33.4$\pm$1.1 &    4.4 &  1.1 \\
AC 12                &    197 &   10.9$\pm$0.7 &    3.7 &  1.6 \\
\hline
\end{tabular}
\end{table}

Table~\ref{tab:otherfeats} provides an inventory of other features
recognized in the EFIGI galaxies. A barlens (bl) is a recently
recognized feature (Laurikainen et al. 2011) that is intermediate in
scale between a nuclear lens (nl) and an inner lens (l), relative to
the bar length. Generally, in galaxy morphology, a lens is a feature
with a shallow brightness gradient interior to a sharper edge (section
8.9).  Barlenses are most obvious in strong bar cases, and have been
closely linked to ``boxy/peanut" bulges (Laurikainen and Salo 2017;
Salo  and Laurikainen 2017). Table~\ref{tab:catalog} includes 272
recognized cases of a barlens, or 6.1\% of the full EFIGI sample. The
average stage of barlens galaxies in Table~\ref{tab:catalog} is S\uab,
and they are found mainly in the stage range S\u0a\ to S\ubc.

Nuclear rings (nr) and related features, such as nuclear pseudorings
(nr$^{\prime}$), nuclear ring-lenses (nrl), and nuclear lenses (nl) are
recognized in only 102 of the sample galaxies. These occur at an
average stage of S\aub\ over the stage range S0/a to S\buc. While the
mean family of barlens galaxies is \SAuB, that for nuclear rings in the
sample is \SuAB. This subtle distinction may be a selection effect.
Comer\'on et al. (2010) showed that nuclear rings tend to be smaller in
more strongly-barred galaxies, making them less likely to be
recognized.  Nuclear bars (nb) and nuclear spirals (ns) tend to be
phenomena of later stages, on average: Sb for nuclear bars and
S\ubc\ for nuclear spirals. Resolution effects in detecting nuclear
features are discussed by Buta (2017a).

Table~\ref{tab:otherfeats} also includes the mean stages, families, and
ranges for X patterns and ansae-type bars. A galaxy classified as
SB$_x$ (or SAB$_x$) may either be a case of an edge-on boxy peanut
bulge or a lower inclination galaxy having a strong bar with an inner
boxy character (section 8.3). A galaxy where the bar has a bright, rounder inner
section flanked by two enhancements (``ansae") is called an ansae-type
bar (deVA; Sandage 1961), and is classified as type SB$_a$ (or SAB$_a$).
Some galaxies show both ansae and an inner boxiness; these are
classified as SB$_{xa}$ or SB$_{ax}$. In Table~\ref{tab:catalog}, B$_x$
and B$_a$ galaxies both average being found in S\uab\ galaxies, while the
B$_{xa}$ and B$_{ax}$ cases average at stage S\0ua.

Spindle (or highly-inclined) galaxies (sp) constitute 23\% of the full
EFIGI sample. An additional 258 (5.8\%) of the sample are classified as
warped spindles (spw). Spindles and warped spindles average near stage
Sc.

Finally, the classifications of 665 EFIGI galaxies (14.9\% of the
sample) are appended with ``pec", indicating a peculiar feature, often
an asymmetry or peculiar dust pattern.

\begin{table*}
\centering
\caption{Other Features
}
\label{tab:otherfeats}
\begin{tabular}{llclcrr}
\hline
Feature & $<Stage>$  & Range  & $<Family>$ & Range & $n$ & \%$N$=4458     \\
1             &     2     &     3     & 4 & 5 & 6  & 7  \\
\hline
bl & S\uab         & S\u0a         to S\ubc         & \SAuB & \SAuB\ to SB    &   272 & 6.1 \\
nr,nr$^{\prime}$,nrl,nl         & S\aub         & S0/a          to S\buc         & \SuAB & SA    to \SAuB &   102 & 2.3 \\
nb         & Sb            & Sa            to S\ucd         & SAB   & \SuAB\ to \SAuB &    43 & 1.0\\
ns         & S\ubc         & Sa            to Scd           & \SAuB & SAB   to SB    &    13 & 0.3 \\
B$_x$         & S\uab         & S\u0a         to Sbc           & SAB   & \SuAB\ to SB    &    92 & 2.1\\
B$_a$       & S\uab         & S0$^+$        to Sbc           & \SAuB & SAB   to SB    &   218 & 4.9 \\
B$_{xa}$,B$_{ax}$    & S\0ua         & S0$^+$        to Sab           & \SAuB\ & SAB   to SB    &    44 & 1.0 \\
sp       & S\buc         & Sa            to Sdm           & ..... &  .....   &  1032 & 23.1 \\
spw         & Sc            & S\aub\ to S\udm         & ..... &  .....   &   258 & 5.8 \\
pec        & Sb            & S0$^+$        to Sd            & ..... & ..... &   665 & 14.9 \\

\hline
\end{tabular}
\end{table*}

\section{Morphological Highlights of the Catalogue}

In this section, attention is brought to exceptional or special
morphological features found in the EFIGI sample. These have not
necessarily been previously studied in spite of multiple classifiers.

\subsection{Extremely Oval Inner Rings in SB Galaxies}

\begin{figure}
\includegraphics[width=\columnwidth]{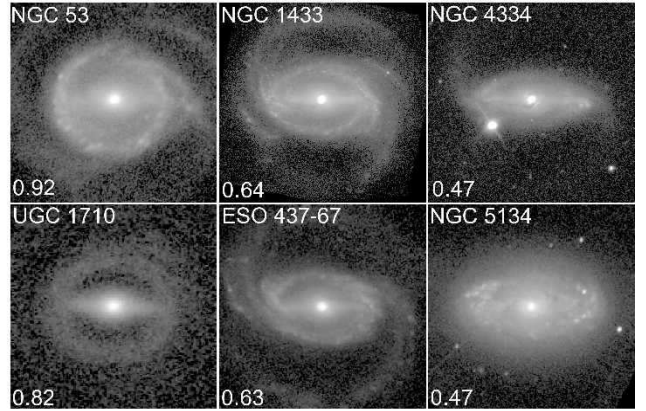}
\caption{Montage showing the wide range of intrinsic shapes of inner
rings in SB galaxies, based on deprojected $B$ or $g$-band images
(foreground stars removed in some panels). Only NGC 4334 and UGC 1710
are in the EFIGI sample; the remainder are from the CSRG.  The ring
face-on axis ratio is given at lower left.}
\label{fig:ngc4334}
\end{figure}

The typical inner ring in an SB galaxy has an intrinsic axis ratio of
0.81$\pm$0.06 (Buta 1995). An extremely oval inner ring is one which
has a much smaller intrinsic axis ratio than this.  In the EFIGI
database, NGC 4334 (Figure~\ref{fig:ngc4334}, upper right) is an
interesting example where the intrinsic axis ratio of the inner ring is
$\approx$0.5. Such extremely oval inner rings are rare but are
important for what they highlight about barred galaxy dynamics and star
formation. The distribution of star formation around inner rings is
very sensitive to ring shape (Crocker et al. 1996; Grouchy et al.
2010). Figure~\ref{fig:ngc4334} shows the wide range of intrinsic
shapes of inner rings that any theory would have to explain, based on
two examples from Table~\ref{tab:catalog} and four examples from the
Catalogue of Southern Ringed Galaxies (CSRG; Buta 1995). Other examples
are illustrated by Buta (2017a).

\subsection{Young and old stellar population bar ansae}

A large sample of color images as provided by the SDSS allows us to see
certain phenomena in a different light. One such phenomenon is bar
ansae, which we mentioned in section 7. These features are common among
early-type barred galaxes (Martinez-Valpuesta et al.  2007), and even
inspired an early theoretical study (Danby 1965). In the discussion by
Martinez-Valpuesta et al. (2007), it was noted that most ansae appear
stellar dynamical in nature, although at least one example of
star-forming ansae (in NGC 4151) was presented. Ansae have recently
been shown using numerical simulations to form in the disk-shaped
remnant of the merger of two spiral galaxies (Athanassoula et al.
2016).

\begin{figure}
\includegraphics[width=\columnwidth]{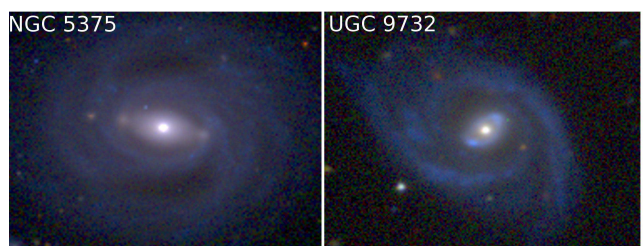}
\caption{Two intermediate-type spirals showing contrasting bar ansae
colors.
}
\label{fig:ansae}
\end{figure}

EFIGI provides many examples that can change our view of these
features. Figure~\ref{fig:ansae} shows two contrasting cases: NGC 5375,
an S\ubc -type spiral with old stellar population ansae, and UGC 9732,
an S\buc -type spiral with strong blue ansae.  Old stellar-population
ansae are of three types: linear, short arcs, and circular spots, and
are generally seen in early-to-intermediate type spirals. In contrast,
blue ansae seem to avoid early-type galaxies. Linear ansae can account
for the strong boxy shapes of the ends of some bars (Athanassoula et
al. 1990). Blue ansae could be linked to extremely oval inner rings.

\subsection{X galaxies and Box/Peanut Bulges}

X patterns in galaxies were first recognized by Whitmore \& Bell
(1988), who suggested the strange features are caused by accretion of a
companion. However, as reviewed by Athanassoula (2005), box/peanut
bulges are nothing more than side-on views of bars. X patterns may be
an optical illusion and are believed to show the vertical resonant
structure of bars (Bureau \& Freeman 1999; Athanassoula \& Bureau 1999;
Bureau et al. 2004, 2006).

\begin{figure}
\includegraphics[width=\columnwidth]{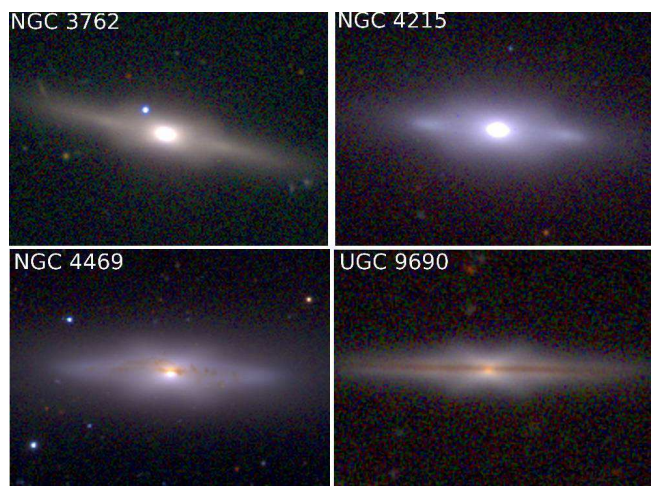}
\caption{Four almost exactly edge-on EFIGI S0 and S galaxies showing strong
X and box/peanut bulge patterns}
\label{fig:Xgals}
\end{figure}

The EFIGI database includes many excellent examples of X patterns and
box/peanut bulges. Four conventional edge-on examples are shown in
Figure~\ref{fig:Xgals}. In non-edge-on early-type galaxies, an X
pattern is often seen in an inner boxy zone that is generally flanked
by ansae.  In Figure~\ref{fig:Xgals}, NGC 4215 has two very strong
ansae flanking a very diffuse box/peanut zone. This case, more than any
other, suggests that ansae are much flatter than the inner zones of
ansae bars.

\subsection{Hexagonal Zone X Galaxies}

\begin{figure}
\includegraphics[width=\columnwidth]{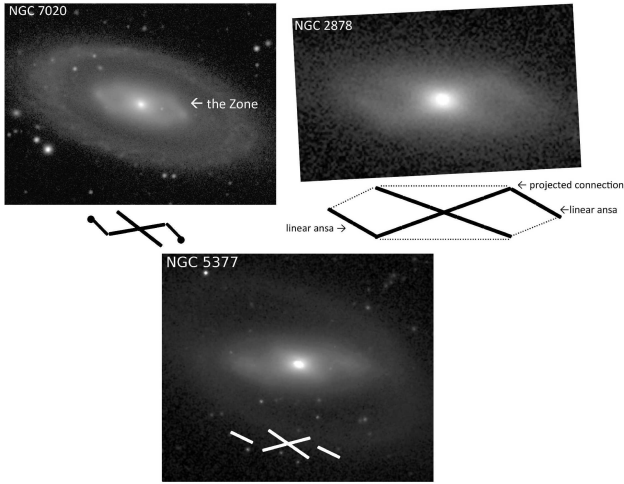}
\caption{{\it top left}: $B$-band image of NGC 7020, a southern galaxy
showing a bright outer ring surrounding an unusual hexgonal zone
crossed by a subtle X pattern (Buta 1990). {\it top right}:  EFIGI
galaxy NGC 2878 also shows a hexagonal zone X pattern; however, in this
case the pattern represents the whole galaxy, not just the inner
section; {\it bottom}: EFIGI galaxy NGC 5377, showing an inner pattern
similar to NGC 7020 but lacking the hexagonal shape. The schematics
below each image show how the likely projection of a 3D X-pattern
associated with the inner part of a bar projects onto what are likely
flat, linear ansae.}
\label{fig:hexzonesA}
\end{figure}

NGC 7020 (Figure~\ref{fig:hexzonesA},upper left) is an unusual non-EFIGI
early-type southern galaxy with an exceptionally bright and
well-defined outer ring surrounding a distinct inner hexagonal zone
that appears crossed by a subtle X pattern (Buta 1990). It is the
prototype of what may be called a ``hexagonal zone X galaxy." While X
patterns are most commonly seen in nearly edge-on galaxies and are
thought to be related to bars, NGC 7020 is inclined only 69$^o$ (deVA).
The EFIGI sample includes at least 15 cases that resemble NGC 7020 in
various ways.

Hexagonal zone X galaxies are easily explained in terms of an inclined
view of a bar having a 3D inner section flanked by two flat, almost
linear ansae. The X arises from vertical resonant bar orbits (e.g.,
Athanassoula 2005). Figure~\ref{fig:hexzonesA} shows how the hexagonal
zones come about: a galaxy having two linear bar ansae is viewed at an
intermediate angle that allows one set of the ends of the X to project
as connecting to one pair of ends of the ansae. This effect is
especially well seen in EFIGI galaxy NGC 2878 (Figure~\ref{fig:hexzonesA},
upper right), in which the hexagonal zone X morphology is the main
morphological feature. The projection of one arm of the X against the
ends of the ansae gives the likely false impression that NGC 2878 is a
spiral galaxy.

The bottom panel of Figure~\ref{fig:hexzonesA} shows NGC 5377, an inclined
(but not edge-on) early-type EFIGI spiral showing a prominent inner X
pattern and bright, slightly curved ansae (Laurikainen et al. 2011).
Although the inner zone is similar to NGC 7020, the
appearance is not hexagonal. In this case, as shown by the schematic,
the arms of the X do not project onto the ends of the ansae. The
appearance of the inner zone is that of a strongly skewed spiral bar.
In this case, however, the skewness is mostly an artifact of projection
effects. This is consistent with the numerical simulations of
Athanassoula et al. (2015) and most recently with simulations made by
Erwin and Debattista (2013) and Salo and Laurikainen (2017).

\subsection{Skewed Spiral Bars}

Erwin and Debattista (2013) show that the combination of a 3D
box/peanut bulge with much flatter bar ends can cause the bar of a
galaxy to take on a ``box + spurs" look. Such bars can look skewed in
characteristic ways. In these bars, the observed skewness is not
necessarily real, but a projection effect between the 2D and 3D parts
of the bar. However, when spiral-like bars are seen in relatively
face-on galaxies, the skewness is likely to be real.

\begin{figure}
\includegraphics[width=\columnwidth]{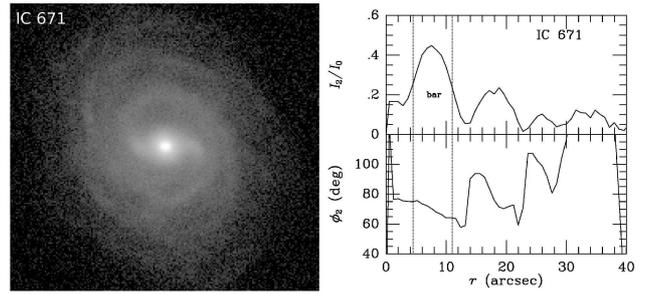}
\caption{The bar in IC 671 is skewed in a trailing sense. The graphs
show the $m$ = 2 relative Fourier amplitudes and phases as a function
of radius. The vertical dotted lines delineate the bar region of the
galaxy.
}
\label{fig:ic0671}
\end{figure}

One of the best cases in the EFIGI sample is IC 671
(Figure~\ref{fig:ic0671}). The galaxy shows a spiral-like bar within a
large inner pseudoring. Outer isophotes favor an inclination of only
37$^o$. Fourier analysis of a deprojected image gives a gaussian
relative $m$=2 intensity profile for the bar (Buta et al.  2006), but
the phase, $\phi_2$, for this bar is not constant with radius
(Figure~\ref{fig:ic0671}, right).

\begin{figure}
\includegraphics[width=\columnwidth]{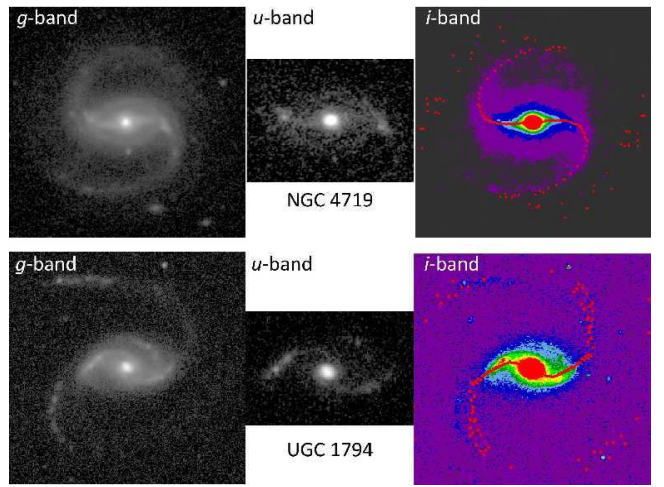}
\caption{Bar skewness in the nearly face-on galaxies UGC 1794 and NGC 4719. The skewness 
is mapped in the right frames using the phase of the $m$=2 Fourier component. The $u$-band
images in both cases show an ansae character in the distribution of star formation. The
estimated inclination of NGC 4719 is 22$^o$ based on ellipse fits to $g$-band outer isophotes.
UGC 1794 has been assumed to be face-on.
}
\label{fig:skew1}
\end{figure}

\begin{figure}
\includegraphics[width=\columnwidth]{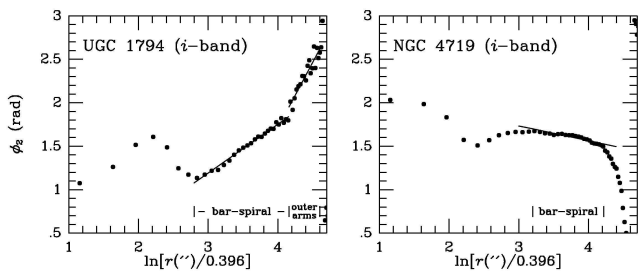}
\caption{Graphs of phase $\phi_2$ versus the natural log of the radius
for UGC 1794 and NGC 4719, useful to estimate the pitch angles of the
bar-spirals. {\it left}: The profile for UGC 1794 shows two radial
zones where $\phi_2$ versus ln$r$ is linear. For the bar-spiral zone,
the pitch angle is 61$^o$, while for the outer arms the pitch angle is
34$^o$.  {\it right}: For NGC 4719, the skewed bar is in the radial
zone indicated. From a linear fit to this zone, the pitch angle of the
bar-spiral is 80$^o$.  The outer arms of this galaxy are partly
distorted towards an R$_1^{\prime}$ outer pseudoring, and as a result
the pitch angle tends to 0$^o$ in the outer disk.
}
\label{fig:skew2}
\end{figure}

Figure~\ref{fig:skew1} shows two even more low inclination systems
showing skewed bars. UGC 1794 is a clear and unambiguous example of a
``spiral bar" (Buta 1986), a spiral distributed within a broad but very
bar-like oval zone.  The ($\phi_2$,ln$r$) profile (Figure~\ref{fig:skew2},
left) for UGC 1794 shows two radial zones where $\phi_2$ versus ln$r$
is linear. For the bar-spiral zone, the pitch angle is 61$^o$, while
for the outer arms the pitch angle is 34$^o$. For NGC 4719, the skewed
bar is in the radial zone indicated in Figure~\ref{fig:skew2}, right. From
a linear fit to this zone, the pitch angle of the bar-spiral is
80$^o$.  The outer arms of this galaxy are partly distorted towards an
R$_1^{\prime}$ outer pseudoring, and as a result the pitch angle tends
to 0$^o$ in the outer disk.

Bar skewness, if real, can drive secular evolution of the stellar mass
distribution through the potential-density phase shift that would
result from the skewed shape (Zhang 1996; Zhang \& Buta 2007).

\subsection{Intrinsic Bar-Ring Misalignment}

Statistics of the projected relative position angle between the long
axis of a bar and the major axis of a ring have indicated that rings
are intrinsically elongated and have preferred alignments with respect
to bars (Buta 1995). The ``rule" for inner SB rings is alignment
parallel to the bar, while the ``rule" for R$_1$ and R$_1^{\prime}$
outer rings is alignment perpendicular to the bar.  In spite of what
statistics showed, however, Buta (1995) nevertheless identified several
nearby face-on galaxies where an inner ring and a bar are significantly
misaligned, clear counter-examples to the normal situation. More
recently, Comer\'on et al. (2014) used statistics of rings in the
S$^4$G (Sheth et al. 2010) to show that, among late-type galaxies,
there is a significant population of inner rings that are
misaligned and even perpendicular to the bars they enclose.

\begin{figure}
\includegraphics[width=\columnwidth]{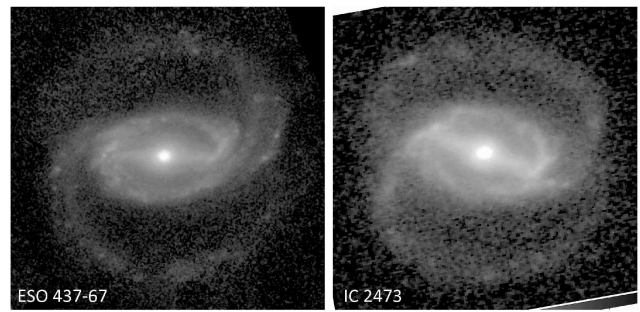}
\caption{Deprojected images showing misaligned bar-R$_1^{\prime}$
galaxy IC 2473 ($g$-band) with a more normal, aligned example ESO
437$-$67 ($B$-band; Buta \& Crocker 1991)
}
\label{fig:ic2473}
\end{figure}

The EFIGI sample has brought attention to bar-outer pseudoring
misalignment. An excellent example (previously noted by Buta 2017b) is
IC 2473, a galaxy showing a strong outer pseudoring having a clear
dimpled, figure-eight shape (Figure~\ref{fig:ic2473}). These are
characteristics of outer resonant subclass outer pseudorings symbolized
in the CVRHS by R$_1^{\prime}$, a type generally aligned perpendicular
to bars (Buta 1995). Yet, when deprojected, Figure~\ref{fig:ic2473} shows
that the R$_1^{\prime}$ ring in IC 2473 is at an intermediate angle
with the bar. It appears that in IC 2473, the inner (r) and outer
(R$_1^{\prime}$) are elongated nearly perpendicularly, and that the bar
is misaligned with both features.  Figure~\ref{fig:ic2473} compares IC 2473
with a deprojected $B$-band image of non-EFIGI galaxy ESO 437$-$67 (Buta \& Crocker
1991), a more typical example where the same kinds of features have the
standard alignments.

Bar-ring misalignment is important because it challenges existing
models of resonance rings and is unlikely to be a stable, long-term
phenomenon. Such misalignment could point to the possibility that the
bar has a different pattern speed from that of the combined
rR$_1^{\prime}$ pattern (e.g., Rautiainen \& Salo 2000).

\subsection{Barred spirals with non-outer pseudoring arms}

\begin{figure}
\includegraphics[width=\columnwidth]{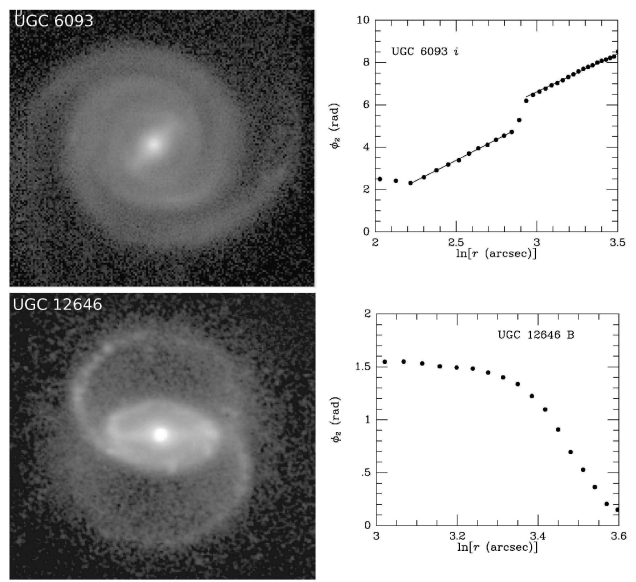}
\caption{Comparison of two barred galaxies, one having logarithmic
spiral structure (UGC 6093, assumed face-on) and the other
non-logarithmic arms (UGC 12646, non-EFIGI sample, deprojected). The
graphs show the Fourier $m$=2 phases of the arm patterns}
\label{fig:ugc6093}
\end{figure}

Spiral patterns in nonbarred galaxies are usually logarithmic and can
be described by a single value of the pitch angle. The spiral structure
of barred galaxies can also be logarithmic, but shows more diversity
through the existence of the outer resonant subclasses of outer rings
and pseudorings. In these cases, the pitch angle of spiral arms is not
constant with radius and the patterns close in characteristic ways, one
of which is shown in Figure~\ref{fig:ugc6093}.  The existence of such
dynamically identifiable patterns has been attributed to secular
evolution of more open spiral patterns and the role of pattern speeds
on galaxy morphology (Buta \& Combes 1996). While outer resonant
subclass rings might be bar-driven patterns having the same pattern
speed as the bar, it is likely that a logarithmic spiral in a barred
galaxy has a pattern speed different from the bar and may be an
independent pattern, not driven by the bar.

\subsection{Extreme Late-type Barred Spirals}

In Table~\ref{tab:catalog}, SB(s)\cud\ is a common morphology involving
extreme late-type galaxies with conspicuous bars. What makes these
galaxies important is threefold: (1) they are essentially pure disk
galaxies, a type that has been difficult to produce in cold dark matter
simulations like the $\Lambda$CDM model (e.g., Mayer et al. 2008; see
also Robertson et al.  2004); (2) the bars of these galaxies are very
different from the typical bar seen in earlier types in that they lack
a broad inner component, are probably vertically very thin and
azimuthally relatively narrow, and are dominated by a younger stellar
population; and (3) these galaxies are part of the tendency for the bar
fraction to rise among late-type galaxies (e.g., Barazza et al. 2008).

\begin{figure}
\includegraphics[width=\columnwidth]{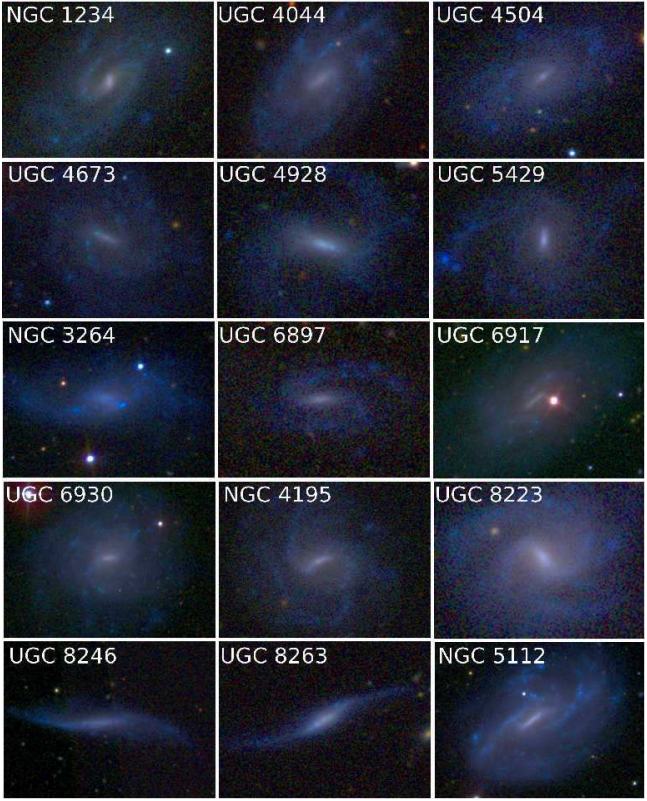}
\caption{Fifteen EFIGI sample galaxies classified as (or close to)
type SB(s)\cud\ in Table~\ref{tab:catalog}.
}
\label{fig:sbsd}
\end{figure}

Figure~\ref{fig:sbsd} shows 15 EFIGI sample galaxies classified as (or
close to) type SB(s)\cud\ in Table~\ref{tab:catalog}. Apart
from a somewhat irregular appearance, such galaxies look remarkably
homogeneous with respect to bar and disk colors. More detailed study of
such galaxies could shed important light on the nature of bars in a
pure disk habitat.

\subsection{Ringed versus lensed galaxies}

Rings and lenses in nonbarred galaxies are of special interest because
the formation of both types of features is thought to be related to
bars. A ring can form by gas accumulation at resonances, under the
continuous action of gravity torques from a bar (Buta \& Combes 1996).
The origin of lenses is under debate. Kormendy (1979) argued that an
inner lens may form from dissolution of a bar, owing to an interaction
between a bar and the other components in a galaxy (see also Bournaud
\& Combes 2002). Alternatively, a lens may form from dissolution of a
``dead" ring no longer forming any new stars (e. g., NGC 7702; Buta
1991). In a recent numerical study, Eliche-Moral et al. (2018) show
that major mergers can account for much of the inner structure of S0
galaxies, including rings, ovals, lenses, and inner discs.

Although most rings do appear to be related to bars, the number of
nonbarred galaxies with these features is not negligible. In fact, some
of the most spectacular examples of star-forming rings are found in
mostly nonbarred galaxies [e.g., NGC 4736 (Schommer \& Sullivan 1976;
deVA) and NGC 1553 (Kormendy 1984); see also Crocker et al. 1996 and
Grouchy et al. 2010]. The weak link between rings and bars is further
examined by D\'iaz-Garc\'ia et al. (2019).

\begin{figure}
\includegraphics[width=\columnwidth]{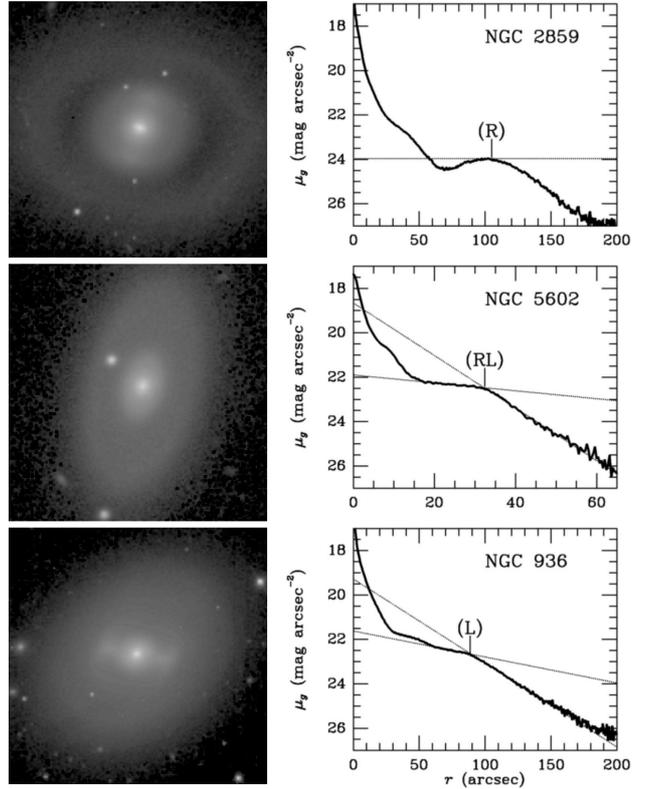}
\caption{Mean major axis $g$-band luminosity profiles (right panels) of
three EFIGI galaxies showing the distinction between outer rings (R),
outer ring-lenses (RL), and outer lenses (L).  The left panels show
SDSS $g$-band images of the same galaxies in units of mag
arcsec$^{-2}$. The lines for NGC 5602 and NGC 936 are fits to the inner
and outer parts of the lens profiles.}
\label{fig:ringslenses}
\end{figure}

The EFIGI sample includes some interesting examples of ringed and
lensed galaxies. Figure~\ref{fig:ringslenses} shows how the
classifications (R), (RL), and (L) (outer ring, outer ring-lens, and
outer lens, respectively) translate into mean major axis $g$-band
surface brightness profiles for NGC 2859 (\SAuB ), NGC 5602 (SA), and
NGC 936 (SB). Although the outer feature of NGC 5602 is subtly
ring-like in the SDSS $g$-band image, the major axis profile shows
no clear enhancement, while the outer ring of NGC 2859 is a strong
enhancement. The outer lens of NGC 936 is strongly sloped near its
``edge" (intersection of the two lines).

\begin{figure}
\includegraphics[width=\columnwidth]{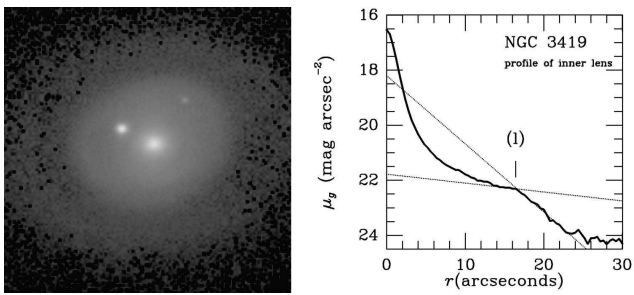}
\caption{SDSS $g$-band image in units of mag arcsec$^{-2}$ (left) and
mean major axis $g$-band luminosity profile (right) of NGC 3419, a
strong inner lens galaxy. The lines are fits to the inner and outer
parts of the lens profile.
}
\label{fig:ngc3419}
\end{figure}

NGC 3419 is a mostly SA galaxy which has a very bright inner lens (l)
and a faint but distinct outer ring (Figure~\ref{fig:ngc3419}, left).
The $g$-band mean surface brightness profile along the (l) major axis
(Figure~\ref{fig:ngc3419}, right) provides a good example of the
definition of a lens as a feature with a shallow brightness gradient
interior to a sharp edge (Kormendy 1979).

\section{The CVRHS system and Large-Scale Automated Galaxy Classification}

Although the CVRHS system is a fairly accessible approach to galaxy
classification, it may not be practical for it to be ``crowd-sourced"
in the manner of the Galaxy Zoo project, nor is it likely that the
system will ever be directly applied to more than a few tens of
thousands of galaxies. Yet, as noted by Dom\'inguez S\'anchez et al.
(2019), astronomy is entering a new era of large surveys (e.g., Euclid,
LSST, WFIRST, etc.) that will require the most sophisticated tools of
automated classification possible to handle the literally millions of
galaxies that will need to be classified.

A general requirement for automatic classification is a fairly large
training set, i.e., a set of images of galaxies whose classifications
are already known, either by expert classification or by the
crowd-sourcing approach of Galaxy Zoo 2, for example. Dom\'inguez
S\'anchez et al. (2018) use the detailed classifications of Nair and
Abraham (2010) and Galaxy Zoo 2 to train a deep machine learning
algorithm based on convolutional neural networks (Dieleman et al. 2015)
to automatically classify 670,000 galaxies. In principle, the
classifications in Table 4 could be used as such a training set,
especially with regard to the numerical T-types and F-types. However,
the complexities of inner, outer, and nuclear varieties, as well as
other aspects of CVRHS morphology, may challenge automatic recognition.

The advantage of CVRHS morphology lies not only in its usefulness for
training an automated classification algorithm, but also in recognizing
peculiar or special cases of interest. This is true of all visual
surveys, including Galaxy Zoo 2.

\section{Summary}

The EFIGI galaxy sample (Baillard et al. 2011) has been re-examined
from the point of view of CVRHS classification. The main results from
this study are:

\noindent
1. a consistent set of detailed classifications of RC3 galaxies that
likely represent an improvement on the many RC3 classifications that
were based on small-scale sky survey prints. The CVRHS types have
internal dispersions of $\sigma_i(T)$ = 0.7 stage intervals,
$\sigma_i(F)$ = 0.6 family intervals, and $\sigma_i(IV)$ = 1.1 inner
variety intervals.

\noindent
2. good agreement between the mean stage and family classifications in
Table 1 and those of Baillard et al. (2011), RC3, Nair \& Abraham
(2010), and Ann et al. (2015). The external dispersions in the CVRHS
classifications are $\sigma_e(T)$ = 1.1 stage intervals and
$\sigma_e(F)$ = 0.8 family intervals.

\noindent
3. a bar fraction of 53\%-67\%, which is typical of a
non-volume-limited sample like EFIGI. The peculiar feature of CVRHS bar
classifications, which in the S$^4$G mid-IR classifications of Buta et
al. (2015) show a prominent minimum in bar fraction around stages Sbc
to Sc, reappears in the classifications of the EFIGI galaxies. This is
not necessarily a personal equation effect, but may show that the stage
sequence for barred galaxies is not necessarily in step with that for
nonbarred galaxies.

\noindent
4. the highest relative frequency of inner rings and pseudorings, as
well as the highest frequency of outer features, occurs near stage Sab.
This is consistent with previous results and is one of the most
important observations along the Hubble sequence.

\noindent
5. the highest relative frequency of inner rings and pseudorings
occurs in SB galaxies; however, the highest frquency of inner ring-lenses
and lenses occurs in SA galaxies.

\noindent
6. the sample emphasizes grand-design, multi-armed spirals in the type range
Sab to Sbc; flocculent spirals in the sample average between S\ucd\ to Sdm.

\noindent
7. the VRHS classification volume has a more asymmetric shape than
usually depicted because of nonuniform fillings of the ``cells" from
earlier to later stages.

\noindent
8. further verification of the wide range, $\approx$0.5-1.0, of the
 intrinsic face-on minor to major axis ratios of inner rings.

\noindent
9. further recognition of the close relation between ansae bars and X
patterns/box-peanut bulges. 

\noindent
10. recognition of bar-outer pseudoring misalignment among outer resonant subclass
rings.

\noindent
11. recognition of genuine bar skewness that in some galaxies could drive secular
evolution.

\noindent
12. identification of unusual barred galaxies having logarithmic outer
spiral patterns rather than outer pseudorings

\noindent
13. many new examples of bar ansae, particularly the previously
under-appreciated class of blue ansae and the extension of ansae
into later type galaxies.

\noindent
14. an established connection between ``hexagonal zone X galaxies"
and projection effects in inclined bars with a 3D inner section.

\noindent
15. finally, many excellent examples of different galaxy types as
per the construction of the EFIGI sample by Baillard et al. (2011).

The author thanks the referee for helpful comments which improved this
paper. This work was supported initially by a grant from the Research
Grants Committee of the University of Alabama. Funding for the SDSS and
SDSS-II has been provided by the Alfred P. Sloan Foundation, the
Participating Institutions, the National Science Foundation, the U.S.
Department of Energy, the National Aeronautics and Space
Adiminstration, the Japanese Monbukagakusho, the Max Planck Society,
and the Higher Education Funding Council for England. The SDSS website
is http://www.sdss.org.

\end{document}